\documentstyle[12pt,epsf]{article}
\begin{document}
\input{FEYNMAN}
\def
\dslash{\partial \!\!\! /}
\def
\dbar{\bar\Delta}

%
%
\def\ap#1#2#3{     {\it Ann. Phys. (NY) }{\bf #1} (19#2) #3}
\def\arnps#1#2#3{  {\it Ann. Rev. Nucl. Part. Sci. }{\bf #1} (19#2) #3}
\def\ijmpa#1#2#3{  {\it Int. J. Mod. Phys. }{\bf A#1} (19#2) #3}
\def\npb#1#2#3{    {\it Nucl. Phys. }{\bf B#1} (19#2) #3}
\def\plb#1#2#3{    {\it Phys. Lett. }{\bf B#1} (19#2) #3}
\def\prd#1#2#3{    {\it Phys. Rev. }{\bf D#1} (19#2) #3}
\def\prb#1#2#3{    {\it Phys. Rev. }{\bf B#1} (19#2) #3}
\def\pro#1#2#3{    {\it Phys. Rev. }{\bf #1} (19#2) #3}
\def\prep#1#2#3{   {\it Phys. Rep. }{\bf #1} (19#2) #3}
\def\pha#1#2#3{    {\it Physica }{\bf A#1} (19#2) #3}
\def\prl#1#2#3{    {\it Phys. Rev. Lett. }{\bf #1} (19#2) #3}
\def\ptp#1#2#3{    {\it Prog. Theor. Phys. }{\bf #1} (19#2) #3}
\def\rmp#1#2#3{    {\it Rev. Mod. Phys. }{\bf #1} (19#2) #3}
\def\zpc#1#2#3{    {\it Z. Physik }{\bf C#1} (19#2) #3}
\def\mpla#1#2#3{   {\it Mod. Phys. Lett. }{\bf A#1} (19#2) #3}
\def\sjnp#1#2#3{   {\it Sov. J. Nucl. Phys. }{\bf #1} (19#2) #3}
\def\yf#1#2#3{     {\it Yad. Fiz. }{\bf #1} (19#2) #3}
\def\nc#1#2#3{     {\it Nuovo Cim. }{\bf #1} (19#2) #3}
\def\jetp#1#2#3{   {\it Sov. Phys. JETP }{\bf #1} (19#2) #3}
\def\jetpl#1#2#3{  {\it JETP Lett. }{\bf #1} (19#2) #3}
\def\ibid#1#2#3{   {\it ibid. }{\bf #1} (19#2) #3}
\def\ijmpa#1#2#3{  {\it Int. J. Mod. Phys. }{\bf A#1} (19#2) #3}
\def\el#1#2#3{     {\it Europhys. Lett. }{\bf #1} (19#2) #3}
\def\jmmm#1#2#3{   {\it J. Magn. Magn. Mater. }{\bf #1} (19#2) #3}
\def\jpa#1#2#3{     {\it J. Phys. }{\bf A#1} (19#2) #3}

June 14, 1995\hspace{7cm} IP-ASTP-15
\vspace{50pt}
\begin{center}
{\large\sc{\bf Renormalization Group transformations of the decimation type in more than one dimension.}}

\baselineskip=12pt
\vspace{20pt}
 
V. Kushnir\footnote{Email: phvk@phys.sinica.edu.tw.\\
Permanent address: Institute for Theoretical Physics, Metrologicheskaya st.14b, 252143, Kiev-143, Ukraine, Email: eppaitp@gluk.apc.org} and B. Rosenstein\footnote{Email: baruch@phys.sinica.edu.tw}\footnote{Work was supported by  
National Science Council of R.O.C.}\\[5mm]

Institute of Physics, Academia Sinica, Taipei, 11529, Taiwan, R.O.C

\vspace{20pt}
\end{center}
\begin{abstract}
We develop a formalism for
performing real space renormalization group transformations
of the "decimation type"
using low temperature perturbation theory. This type of transformations beyond
$d=1$ is highly nontrivial even for free theories. We construct such a solution in arbitrary dimensions and develop a weak coupling perturbation theory for it. The method utilizes Schur formula to convert summation over decorated lattice into summation over either original lattice or sublattice. We check the formalism on solvable case of $O(N)$ symmetric Heisenberg chain.
 The transformation is particularly useful
to study models undergoing phase transition at zero temperature
(various $d=1$ and $d=2$ spin models, $d=2$ fermionic models, $d=3,4$
nonabelian gauge models...) for which the weak coupling perturbation theory
is a good approximation for sufficiently
small lattice spacing. Results for one class
of such spin systems, the d=2 O(N) symmetric
spin models ($N\ge 3$)
for decimation with scale factor $\eta=2$ (when quarter of the points is left) are given as an example.
\end{abstract}
\indent PACS number(s):  05.50.+q
\vspace{30pt}

\pagebreak
\vspace{1cm}
\section{Introduction}

The renormalization group (RG) transformations is one of the most powerful
and frequently used conceptual as well as practical tools
in statistical physics and quantum field theory. 
While conceptually the idea of
combining variables on neighbouring sites into complexes is
very simple, in practise it almost always turns out to be rather 
complicated.

Historically, the usefulness of RG transformations was
realized after the d=2 Ising model on triangular lattice
was very elegantly solved by Niemeijer and Van Leeuwen \cite{Niemeijer-Leeuwen} using block spinning.
The nonlinear RG method they used was peculiar to that particular system
and didn't allow generalization to more complicated cases. 
For more general systems Wilson \cite{Wilson}
proposed to use the weak coupling (low temperature)
perturbation theory in momentum space. This was first applied to
scalar $\phi^4$ models and subsequently to spin systems \cite{Nelson}
and lattice gauge theories for thinning by factor 2 \cite{Mutter}.
For complicated systems like these with local gauge symmetries
some approximate methods were
 developed like Migdal-Kadanoff \cite{Migdal-Kadanoff}
approximation, variational RG \cite{Patkos}, mean-field RG \cite{Fittipaldi}
or block spinning using Monte Carlo numerical methods \cite{Swendsen,Shenker-Tobochnik}.
However, unlike perturbation theory, these approximations are
uncontrollable in a sense that it is not clear how to estimate errors.

Exact RG transformations are generally not known
(exceptions are decimations in spin chains $d=1$ and mentioned above very
special cases in $d=2$). Moreover after one RG transformation
the resulting action contains generically infinite number of
interaction terms, and therefore one is forced to make an additional approximation
dropping some of them (hopefully the less relevant ones).
RG transformations are especially useful when applied repeatedly.
This requires self similarity of the approximate effective action
and is justified only around fixed points.

Note however that accurate
thinning of the lattice even just by factor $\eta=2$ can greatly facilitate
the study of a model by means of subsequent MC simulation.

There are several types of the RG transformations. The conceptually
simplest one
is the decimation or thinning of degrees of freedom in the configuration space.
Some degrees of freedom located, for example, on sites with at least one
odd coordinate are simply integrated out.

\begin{picture}(40000,27000)(0,-14000)
\drawline\fermion[\N\REG](2000,-5000)[19000]
\drawline\fermion[\N\REG](7000,-5000)[19000]
\drawline\fermion[\N\REG](12000,-5000)[19000]
\drawline\fermion[\N\REG](17000,-5000)[19000]
\drawline\fermion[\N\REG](22000,-5000)[19000]
\drawline\fermion[\E\REG](0,-3000)[24000]
\drawline\fermion[\E\REG](0,2000)[24000]
\drawline\fermion[\E\REG](0,7000)[24000]
\drawline\fermion[\E\REG](0,12000)[24000]
\put(2000,-3000){\circle*{1000}}
\put(12000,-3000){\circle*{1000}}
\put(22000,-3000){\circle*{1000}}
\put(7000,2000){\circle{1000}}
\put(17000,2000){\circle{1000}}
\put(2000,7000){\circle*{1000}}
\put(12000,7000){\circle*{1000}}
\put(22000,7000){\circle*{1000}}
\put(7000,12000){\circle{1000}}
\put(17000,12000){\circle{1000}}
\put(7000,-3000){\circle{1000}}
\put(17000,-3000){\circle{1000}}
\put(2000,2000){\circle{1000}}
\put(12000,2000){\circle{1000}}
\put(22000,2000){\circle{1000}}
\put(7000,7000){\circle{1000}}
\put(17000,7000){\circle{1000}}
\put(2000,12000){\circle{1000}}
\put(12000,12000){\circle{1000}}
\put(22000,12000){\circle{1000}}

\put(-2000,-9000){Fig. 1. Decimation: full circles belong to sublattice ${\cal L}^*$,}
 \put(-2000,-10500){ while empty circles $x \in {\cal D}={\cal L}-{\cal L}^*$ denote the integrated out sites.}

\end{picture}

 Example is given
on Fig. 1 on which spins at empty circle points are integrated out.
\begin{equation}
Z=\sum_{\phi(X):X \in {\cal L}^*}\sum_{\phi(x):x \in {\cal L}-{\bf  L}}e^{-A[\phi(x),\phi(X)]}=\sum_{\phi(X):X \in {\cal L}^*}e^{-A^{dec}[\phi(X)]}
\label{Z}
\end{equation}
Here and in what follows points of
the coarse lattice are denoted by capital letters.
The resulting effective action $A^{dec}$ contains generally 
interactions of any range. Here ${\cal L}$ is the original $d$ dimensional
lattice while ${\cal L}^*$ is a sublattice. Note that remaining
variables are all the old variables. This is not the case for
the so called block spin transformations.
One defines a linear or a nonlinear combination of the variables
on ${\cal L}$, the block spin:
\begin{equation}
{\bf \phi}(X)=f[X,\phi(x)]
\label{<O>}
\end{equation}
For example, for the $O(N)$ classical spins $S^a$ one can define \cite{Shenker-Tobochnik}
\begin{equation}
{\bf S}^a(X)=\sum_{block X} S^a(x)/|\sum_{block X} S^a(x)|
\label{block}
\end{equation}
This combination is highly ambiguous and success of the transformation
critically depends on it's choice. The main
problem is that it is extremely
difficult in practise to perform such a transformation even
perturbatively.
The relations like eq. (\ref{block}) are very nonlinear and even singular \cite{Griffiths}.

Another type of RG transformations, used especially extensively
in field theory, is the momentum space RG \cite{Wilson,Polchinski}.
One defines the momentum space variables
\begin{equation}
\phi(p)\equiv 1/(2 \pi)^d \int d^d x  e^{i p x} \phi(x)
\label{f(p)}
\end{equation}
Now one performs integration over high frequence modes
(strictly speaking
chopping the Brillouin zone, but more often the approximate spherically
symmetric momentum cutoff $\Lambda$ is utilized \cite{Nelson,Ma}). 
This type of RG transformations, while convenient for the $\phi^4$ model, turns out to be especially inconvenient
for constrained systems like the $O(N)$ symmetric spin models.
The reason is following.
While generally in x - space the constraint are local, for example
\begin{equation}
S^a(x) S^a(x)=1
\label{constraint}
\end{equation}
in p - space it becomes a convolution. What does it mean now
high frequency physical modes? The constraint mixes between low and
high frequencies. Since most systems of interest belong to this
class one has to circumvent the difficulty.
One way is to solve the constraint and make the momentum space RG
for physical quantities only. Then
the mode integrated 
effective action contains generally "non-covariant terms". The original
global symmetry is lost since the high frequency modes
do not constitute an $O(N)$ symmetric set. Problems are more
acute with local gauge symmetries.
In practice this type of thinning out of degrees of freedom is often
used for the demonstration purposes only and very rarely the actual
calculations.

Decimation are extremely difficult
to perform even in free theory in more then one dimension (see
for example \cite{Hu}). It might sound surprising that something
is difficult in free theory since all the integrals are Gaussian
and "doable in principle".
Of course it is still a Gaussian integral, but a very complicated one.
Let us consider a free massless boson nearest neighbours action 
\begin{equation}
A[\phi]=-\frac {a^{(d-2)}}{2}\sum_{xy}\phi(x)\Box(x-y)\phi(y)
\label{f(p)}
\end{equation}
where the lattice Laplacian is defined by
$$
\Box(x)\equiv
\sum_{\mu=1}^d \left[\delta(x-\mu)+\delta(x+\mu)-2\delta(x)\right].
$$

If one tries to integrate out a point $\phi(0,...,0)$ the Gaussian integral
involves all its $2d$ nearest neighbours. 
$$\int d \phi(0)\exp\left[ -\frac{a^{(d-2)}}{2}
\left( 2 d \phi(0)^2 -2 \phi(0)\sum_\mu(\phi(\mu)+\phi(-\mu) \right) \right]=
$$
\begin{equation}
=\exp\left[\frac {a^{(d-2)}}{2d} \left(\sum_\mu \phi(\mu)^2+\sum_{\mu\ne\nu}\phi(\mu)\phi(\nu)\right)\right]
\end{equation}
This is very simple.
However when trying subsequently to
integrate another point, say $(2,0...,0)$, all the previous point's neighbours
enter the Gaussian integral and so on. 
The Gaussian integration requires inverting 
increasingly larger
matrices. Since we have to integrate out all the points not belonging to 
the sublattice, some other methods are required. An exception is
the $d=1$ case.
Here the size of the matrix does not grow: integration of a point leads
just to interactions of the neighbouring remaining points.
This is the reason why it is possible in many cases to explicitly find 
decimations in $d=1$.

In this paper we perform the decimation for multidimensional free theories.
The result does not coincides with the naive continuum limit even in the limit
of large $\eta=A/a$. This is discussed in section 2.
Then using this result we develop in section 3 general perturbative
formalism for weakly interacting models.
It is applied in section 4, 5, 6 to the $O(N)$ symmetric nearest neighbours
interaction spin model (the nonlinear $\sigma$-model). In section 4 we derive the general diagrammatic technique for such a models. In section
5 the solvable $d=1$ model is considered and results compared with the usual perturbative ones,
while in 6 the two dimensional asymptotically free model is studied.
The perturbative method generally is better suited for
asymptotically free (=phase transition at $T=0$) models like this one.
This is because on fine lattices coupling becomes small. Then the
RG transformed model can be investigated, say by the MC method,
on coarser lattice. Even simplest decimation with $\eta=2$
reducing the number of points by factor 4 greatly simplifies the numerical work. 

A coarse grained effective action is generally obtained as a series in 
"closeness" of the interacting spins: the nearest neighbours, next 
to nearest etc. In order to make any practical calculations possible, 
one has to truncate it at some point. We restrict our consideration here to the fourth order terms in fields and up to the fourth derivatives.

 We conclude in section 7 by discussing complexity of such calculations and some of their uses.

\section{Decimation of free fields}

Let us start with free boson theory on the lattice. Effective action,
after decimation with parameter $\eta$ generally has a form:
\begin{equation}
A^A[\phi(X)]=\frac{1}{2}\sum_{X\in R^d}\phi(X){\bf \Delta}(X-Y)\phi(Y)
\end{equation}
where bold letters denote sublattice functions.
Of course it is quadratic in $\phi(X)$.

Let us now perform Fourier transforms of the original and sublattice fields
$$
\phi(k)=\frac {a^d}{(2\pi)^{d/2}}\sum_x e^{i a k x} \phi(x)
$$
\begin{equation}
{\bf \phi}(K)=\frac {1}{(2\pi)^{d/2}}\sum_X e^{i K X}
\phi(X).
\end{equation}
This convention fixes the Fourier transforms of propagators:
$$
G(x)= <\phi (x) \phi (0)>=\frac {1}{(2 \pi)^d} \int^{\pi /a}_{-\pi /a} d^dk {\rm e}^{i k x a} G(k),
$$
\begin{equation}
\label{finefourier1}
G(k)=a^d \sum_x {\rm e}^{-i k x a} G(x),
\end{equation}
$$
{\bf G}(X)= <\phi (X) \phi (0)>=\frac {1}{(2 \pi)^d} \int^{\pi}_{-\pi} d^dK {\rm e}^{i K X} {\bf G}(K),
$$
\begin{equation}
\label{coarsefourier1}
{\bf G}(K)=\sum_X e^{-i K X} G(X),
\end{equation}
and inverse propagators
$$
\Delta (x)= \sum_\mu (\delta (x+\mu )+\delta (x-\mu )-2\delta (x))=\frac {a^{(d+1)}}{(2 \pi)^d} \int^{\pi /a}_{-\pi /a} d^dk {\rm e}^{i k x a} \Delta (k),
$$
\begin{equation}
\label{finefourier2}
\Delta (k)=\frac {1}{G(k)}=\frac {1}{a} \sum_x {\rm e}^{-i k x a} \Delta (x)=  \frac{4}{a^2}\sum_\mu {\rm sin}^2 (ak_\mu /2),
\end{equation}
\begin{equation}
\label{coarsefourier2}
{\bf \Delta}(K)=\sum_X e^{i K X} \Delta(X)
\end{equation}
Due to translation invariance, we can invert these in 
momentum space to obtain propagators on the lattice and sublattice correspondingly:
$$
G(x)=\frac {1}{(2\pi)^d}\int_{-\pi/a}^{\pi/a} d^dk e^{- i k x a} \frac {a^2}{4 \sum_\mu {\rm sin}^2 (ak_\mu /2)}
$$
\begin{equation}
{\bf G}(X)=\frac {1}{2\pi}\int_{-\pi}^{\pi}e^{- i K X}\frac {1}{{\bf \Delta}(K)}
\end{equation}

The two models should result in equivalent correlator between
two sublattice points: 0 and $X$:
$G(\eta X)={\bf G}(X)$.
This leads to the following relation between the Fourier transforms:
\begin{equation}
{\bf G}(K)=1/\Delta(K)= \frac {1}{(2\pi)^d}\sum_X e^{-i K X}
\int_{-\pi/a}^{\pi/a} d^dk e^{i k X}
\frac {a^2}{\sum_\mu 4\; {\rm sin}^2 (ak_\mu/2)}
\end{equation}
Summation over $X$ results in sum over $\delta$ functions
\begin{equation}
{\bf G}(K)= \frac {a^d}{(2\pi)^{d}}\int_{-\pi/a}^{\pi/a} d^dk 
\sum\limits_{n=-\infty}^{\infty} \delta ((k-K+2\pi n)_\mu)
\frac {a^2}{\sum_\mu 4\; {\rm sin}^2 (ak_\mu/2)}
\end{equation}
which are used to perform the momentum integrations:
\begin{equation}
G(K)= 
\sum_{n_\mu=1}^{\eta}\frac {a^2}{\sum_\mu 4\; {\rm sin}^2 (a(K+2\pi n)_\mu/2)}
\label{decprop}
\end{equation}
Limits of summation in the last expression follow from the different sizes of Brillouin
 zone for two lattices
(see Fig. 2). Note that due to periodicity the limits of summation 
in $n$ can be shifted by the period $\eta$.

In $d=1$ we recover the previous result since the sum is doable
\cite{Prudnikov}.
In $d>1$, the summation over one of the variables, $n_1$ can be performed similarly,
but the remaining summations should be done numerically.
In particular for

\begin{picture}(40000,23000)(1000,-14000)
\drawline\fermion[\N\REG](2000,-5000)[10000]
\put(-3000,-5000){$K=-\pi$}
\put(-3000,5000){$K=\pi$}
\drawline\fermion[\E\REG](\fermionbackx,\fermionbacky)[40000]
\put(2000,-7000){$k=-\pi /a$}
\put(36000,-7000){$k=\pi /a$}
\drawline\fermion[\S\REG](\fermionbackx,\fermionbacky)[10000]
\drawline\fermion[\W\REG](\fermionbackx,\fermionbacky)[40000]
\drawline\fermion[\NE\REG](2000,-5000)[14000]
\drawline\fermion[\NE\REG](12000,-5000)[14000]
\drawline\fermion[\NE\REG](22000,-5000)[14000]
\drawline\fermion[\SE\REG](15000,-2000)[10000]
\drawarrow[\NW\ATTIP](\fermionfrontx,\fermionfronty)
\put(23000,-9000){$k=K+2 \pi n$}
\drawline\fermion[\NE\REG](32000,-5000)[14000]
\drawline\scalar[\E\REG](2000,0)[18]
\gaplength=700
\drawline\scalar[\N\REG](22000,-5000)[5]
\put(0,-12000){Fig. 2. Brillouin zones for original lattice ($k$) and sublattice ($K$). Lines}
\put(2000,-13500){correspond to $k=K+2 \pi n$.} 
\end{picture}

 $d=2$ we have the propagator:
\begin{equation}
{\bf G}(K)=\frac{1}{\eta}\sum_{n_2 = 1}^{\eta}
   \frac { {\rm sinh} (\eta \, \alpha) {\rm csch} (\alpha ) }
      {2 \left( -1 +
           {\rm \cosh} (\eta \, \alpha) + 2 {\rm sin}^2 (\frac {K_1}{2}) \right) }.
\end{equation}
where
$$
\alpha={\rm arccosh}(1 + 2{\rm sin}^2 (\frac {K_2 + 2\pi n_2}{2\eta}))
$$
For large $\eta$ the Euclidean invariance is restored, ${\bf G}(K_1,K_2)={\bf G}({\sqrt {K_1^2+K_2^2}})$
 and 
numerical calculations show it can be fitted by
\begin{equation}
\label{fit}
G(K)=\frac {1}{K^2} + \frac{1}{2 \pi}\log(\eta)+0.04876 + 0.003022\, K^2 + O[(K^2)^2]
\end{equation}
(see Fig. 3) with an accuracy better 1 percent in all the Brillouin zone. 
The first term is the continuum propagator. Note that the decimated propagator even for large $\eta$
does not coincides with the naive continuum limit.

The contact constant term with logarithmic dependence of $\eta$ is typical for $d=2$ and is nothing else but the bubble integral.
The polynomial coefficients \\[10mm]
are very small and almost coincide with
the Loran

\epsfbox{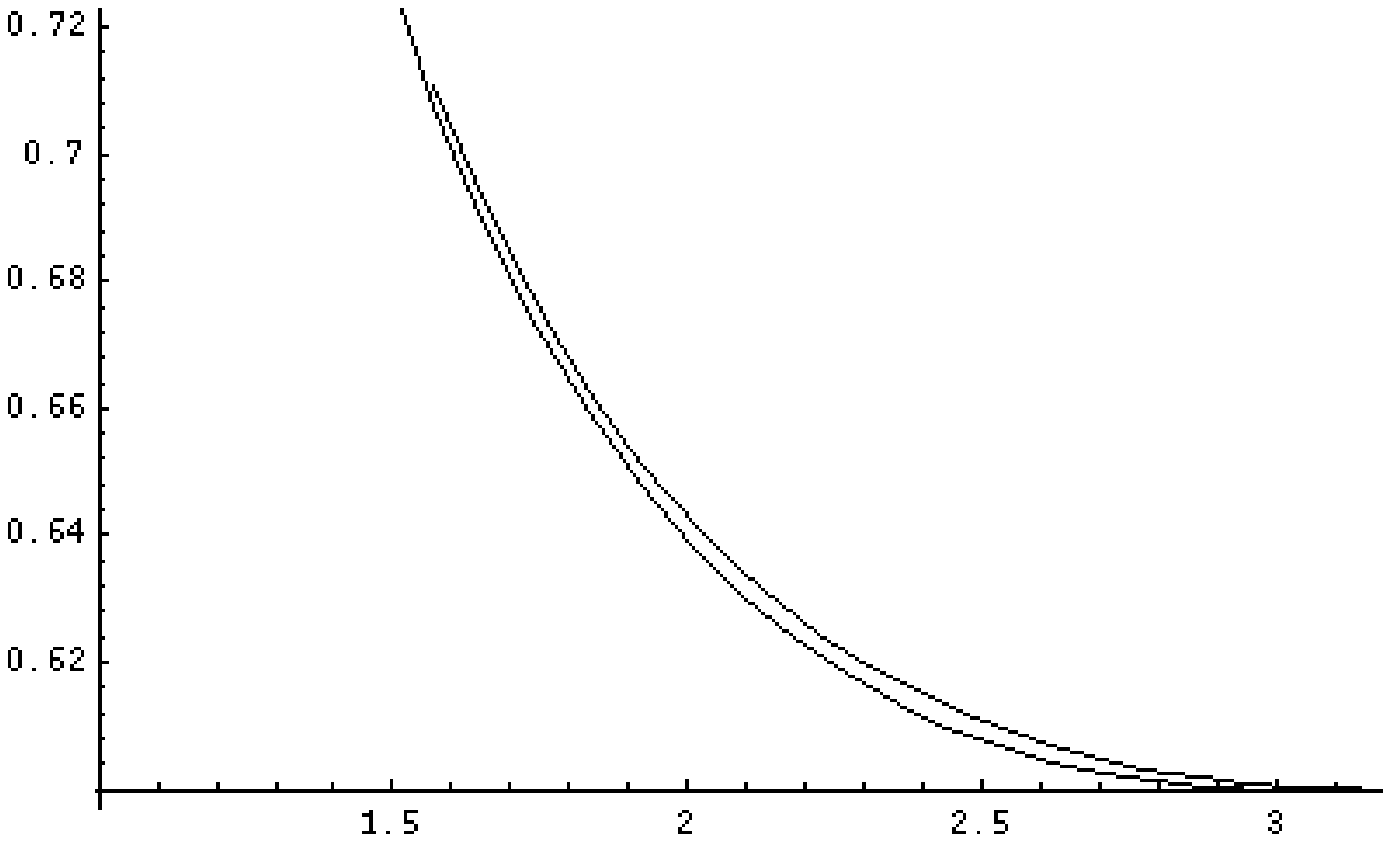}

Fig. 3. $d=2$ free decimated propagator (upper line) and fit eq. (\ref{fit}) (lower line) for $\eta=4$.\\[5mm]
 
 expansion of the propagator around $K=0$.
For finite $\eta$ the symmetry remains of course just the discrete subgroup
of the rotations. 

In higher dimensions similar expressions can be written.
Similar procedure can be extended to free fermion fields (see Appendix A). Especially 
interesting aspect of this is the species doubling \cite{Itzykson-Drouffe}.

In the simplest case of one dimensional massless 
boson field we can explicitly integrate out all the odd points
since the integrals do not intertwine:
\nopagebreak{
$$\prod_{x=-\infty}^{\infty} \int d \phi_{2 x+1}\exp \left[-\frac{1}{2a} \sum_x
\left[ \phi_{2x}^2+ \phi_{2x+1}^2 -(\phi_{2x}+\phi_{2x+2})\phi_{2x+1}\right]\right]
$$
\begin{equation}
=\exp\left[\frac {1}{2 \; (2 a)} \sum_X \left[\phi_{X}^2
+\phi_{X}\phi_{X+1}\right]\right]
\end{equation}}
We therefore obtain the original form with twice lattice spacing: $\beta_2 =\beta /2$, where $\beta$ is an inverse temperature.
The action is a perfect one \cite{Niedermayer}.
For arbitrary $\eta=a/A$ we get similar results, 
$$
\beta_\eta =\beta /2^\eta.
$$

In higher dimensions we still can perform the decimation using momentum space
in the intermediate steps to diagonalize the matrices.

\section{General diagrammatic method for evaluation of the decimated action}

Now we would like to build a perturbation theory for the 
decimation-type RG transformations in the interacting case. For concreteness we discuss the lattice $\phi^4$ 
model
\begin{equation}
A[\phi_x]=a^{(d-2)} \sum_x \left[ \frac{1}{2}(\nabla \phi_x)^2+\frac{m^2}{2}\phi^2_x
+\frac{\lambda}{4!}\phi^4_x \right],
\end{equation}
where $\nabla \phi_x= (\phi_x -\phi_{x-1} )$.
Low temperature (weak coupling)
perturbation theory for this model can be represented via
Feynman diagrams including propagator and the four vertex.
 In momentum space RG, when the high frequency modes from $\Lambda$ to
$\Lambda'=\Lambda/\eta$ are integrated out, 
the resulting effective action on the scale $\Lambda'$ has a general form
\begin{equation}
A^M[\phi_x]=\sum_{n=1}^{\infty} \frac {1}{(2 n)!} \int_{x_1,...x_{2n}} \Gamma^{(2n)}(x_1,..,x_{2n})\phi_{x_1}...\phi_{x_{2n}} 
\end{equation}
The coefficient functions $\Gamma^{(2n)}$ are sums of all the
one particle irreducible contributions with
$2n$ ends. The external momenta are all below $\Lambda'$ while all
the integrated internal momenta are between $\Lambda'$ and $\Lambda$ 
\cite{Wilson}. Since we will significantly modify the
procedure in x space let us briefly outline the p space diagrammatics for
RG.
This is most easily done if original vertices and propagators are split
into several pieces. The vertex decomposes into:
the vertex connecting just high momenta modes (Fig. 4(c)), 
only low momenta modes (Fig. 4(d)) and mixing the two (Fig. 4(e), 4(f), 4(g)). 

\begin{picture}(40000,27000)(0,-17000)
\drawline\fermion[\E\REG](5000,8000)[6000]
\put(8000,6500){a}
\THICKLINES
\drawline\fermion[\E\REG](20000,8000)[6000]
\THINLINES
\put(23000,6500){b}
\drawline\fermion[\E\REG](5000,2000)[6000]
\drawline\fermion[\N\REG](8000,-1000)[6000]
\put(8000,-2500){c}
\THICKLINES
\drawline\fermion[\E\REG](20000,2000)[6000]
\drawline\fermion[\N\REG](23000,-1000)[6000]
\THINLINES
\put(23000,-2500){d}
\drawline\fermion[\E\REG](2000,-6000)[3000]
\drawline\fermion[\N\REG](5000,-9000)[3000]
\put(5000,-10500){e}
\drawline\fermion[\E\REG](5000,-6000)[3000]
\THICKLINES
\drawline\fermion[\N\REG](5000,-6000)[3000]
\THINLINES
\drawline\fermion[\E\REG](13000,-6000)[3000]
\put(16000,-10500){f}
\THICKLINES
\drawline\fermion[\E\REG](\fermionbackx,\fermionbacky)[3000]
\THINLINES
\drawline\fermion[\N\REG](16000,-9000)[3000]
\THICKLINES
\drawline\fermion[\N\REG](\fermionbackx,\fermionbacky)[3000]
\THINLINES
\drawline\fermion[\E\REG](24000,-6000)[3000]
\put(27000,-10500){g}
\THICKLINES
\drawline\fermion[\E\REG](\pbackx,\pbacky)[3000]
\drawline\fermion[\S\REG](\pfrontx,\pfronty)[3000]
\drawline\fermion[\N\REG](\pfrontx,\pfronty)[3000]

\put(0,-13000){Fig. 4. Momentum space RG propagators (a, b) and vertices (c, d, e, f, g)}
\put(0,-14500){for $\phi^4$ model. Low momenta are indicated by bold lines.}

\end{picture}

Propagators are decomposed analogously into low and high frequency parts:
$G(k)=\theta(|k|-\Lambda'){\bf G}(k)+\theta(\Lambda'-|K|)G(K)$.
Note that the coupling between the modes is by means of the vertex
only.
This will be completely different in real space RG.
The integral over high frequencies (denoted $\phi_K$)

$$
{\rm e}^{-A^{eff}[\phi_K]}=
$$
\begin{equation}
 {\rm e}^{-A[\phi_K]} \int_{\phi_k} {\rm exp} \left[-\int_{k>\Lambda'} \left(\frac{1}{2}\phi_k(k^2+m^2)\phi_{-k} \right)-V[\phi_k,\phi_K] \right],
\end{equation}
with
$$
V[\phi_k,\phi_K]
=\frac{\lambda}{4!} \left(\int\limits_{\scriptstyle(|k|,|l|,|m|)>\Lambda'\atop\scriptstyle|k+l+m|>\Lambda'} \phi_k \phi_l \phi_m \phi_{(-k-l-m)} \right.
$$
$$
+\int\limits_{\scriptstyle|K|<\Lambda' ,(|l|,|m|)>\Lambda'\atop\scriptstyle|k+l+m|>\Lambda'} \phi_K\phi_l \phi_m \phi_{(-K-l-m)}+\int\limits_{\scriptstyle(|K|,|L|)<\Lambda',|m|>\Lambda'\atop\scriptstyle|k+l+m|>\Lambda'} \phi_K\phi_L\phi_m\phi_{(-K-L-m)}
$$
\begin{equation}
\left.+\int\limits_{\scriptstyle(|K|,|L|,|M|)<\Lambda'\atop\scriptstyle|k+l+m|>\Lambda'} \phi_K\phi_L\phi_M\phi_{(- k-l-m)}+ {\rm permutations}\right) ,
\end{equation}
 apart from "classical" parts independent of $\phi_k$ is
exponent of the vacuum energy
of the high frequency theory with $\phi_K$ playing a role of the
external sources. This is the sum of all the vacuum diagrams in this
theory. However, as we remarked before the lines do not connect
low to high frequency modes and consequently all the one
particle reducible diagrams vanish.

For the real space RG the perturbation theory can be built in a similar way.
The fields $\phi_X$, $X\in {\cal L}^*$ will be
treated as "external sources", while all the internal points will belong 
to ${\cal D}\equiv {\cal L}-{\cal L}^*$. 
 For $\phi^4$ model this means the following decomposition (Fig. 5). Action is divided into three parts:

\begin{picture}(40000,20000)(0,-10000)
\put(500,8000){\circle*{1000}}
\put(500,6000){a}
\drawline\fermion[\E\REG](10000,8000)[6000]
\put(13000,6000){b}
\put(9500,8000){\circle*{1000}}
\put(16500,8000){\circle{1000}}
\drawline\fermion[\E\REG](23000,8000)[6000]
\put(22500,8000){\circle{1000}}
\put(29500,8000){\circle{1000}}
\put(26000,6000){c}
\drawline\fermion[\E\REG](3500,-1000)[3000]
\drawline\fermion[\E\REG](7500,-1000)[3000]
\drawline\fermion[\N\REG](7000,-4500)[3000]
\drawline\fermion[\N\REG](7000,-500)[3000]
\put(7000,-1000){\circle*{1000}}
\put(7000,-6000){d}
\drawline\fermion[\E\REG](18500,-1000)[3000]
\drawline\fermion[\E\REG](22500,-1000)[3000]
\drawline\fermion[\N\REG](22000,-4500)[3000]
\drawline\fermion[\N\REG](22000,-500)[3000]
\put(22000,-1000){\circle{1000}}
\put(22000,-6000){e}

\put(0,-8000){Fig. 5. Real space RG (decimation) propagators
 (a, b, c) and vertices (d, e)}
\put(0,-9500){for $\phi^4$ model. Full circles belong to sublattice (external fields),}
\put(0,-11000){while empty denote "internal fields" ($x\in {\cal L}-{\cal L}^*$).}

\end{picture}
 "classical action" of 
the "external" field $\phi_X$ (Fig. 5(a), 5(d)),
$$
A^{cl}[\phi_X]=\sum_X\left( \frac {a^{d} m^2}{2} \phi^2_X+ a^{d-2} d\; \phi^2_X+ \frac {a^{d} \lambda}{4!}\phi^4_X\right)
$$
\begin{equation}
\equiv \frac {a^{(d-2)}}{2}\sum_{X,Y} \phi_X A_{XY}\phi_Y+A^{cl}_{int}[\phi_X],
\label{ext} 
\end{equation}
cross-term (Fig. 5(b)) 
\begin{equation}
A_1[\phi_x,\phi_X]=-a^{(d-2)}\sum_{x ,X} \phi_X \dbar(X,x) \phi_x\equiv -a^{(d-2)}\sum_{x ,X} \phi_x B_{x X} \phi_X
\end{equation}
with "external legs"
\begin{equation}
\dbar(X,x)= \sum_\mu(\delta_{X-x+\mu}+\delta_{X-x-\mu})
\label{cross}
\end{equation}
 and an internal part for which all the vertices belong to ${\cal D}\equiv
 {\cal L}-{\cal L}^*$ ("decorated" model) (Fig. 5(c), 5(e)):
\begin{equation}
A_2[\phi_x]=-\frac {a^{(d-2)}}{2} \sum_{x,y} \phi_x D(x,y) \phi_y+\frac {a^{d-2} \lambda}{4!}\sum_x \phi^4_x.
\label{int}
\end{equation}
 Note that unlike momentum space RG, here external fields $\phi_X$ are coupled to internal part only via derivative couplings like off-diagonal part of propagator (Fig. 5(b)) or derivative interaction in nonlinear $\sigma$-model (see the next Sections). All the local vertices will completely decouple into internal (Fig. 5(e)) and external (Fig. 5(d)). 

Integration out all the fields $\phi_x$ will lead to an effective action for the fields $\phi_X$ on sublattice of the form:
\begin{equation}
A[\phi_X]=\sum_{X_1,..,X_{2n}} \frac {1}{(2 n)!} H^{(2n)}(X_1,..,X_{2n})\phi_{X_1}...\phi_{X_{2n}},
\label{acteff}
\end{equation}
where the coefficient functions $H^{(2n)}(X_1,..,X_{2n})$, contrary the momentum space RG, are sums of all the connected contributions with $2n$ ends. These connected functions does not degenerate into one particle irreducible.

To integrate over field $\phi_x$ perturbatively, we need to find its propagator which is matrix inverse to $D$. At the same time, all we know explicitly is the full Laplacian $\Delta$ and original propagator $G=\Delta^{-1}$. Moreover, since ${\cal D}$ does not constitute a sublattice, it is impossible
to make use of Fourier analysis on it. Therefore it is useful to
represent all the summations over ${\cal D}$ via summations over 
${\cal L}$ and ${\cal L}^*$.
This can be done using following algebraic trick \footnote{Somewhat similar technique based on the Schur formula was developed for quasiperiodic systems on the octagonal lattice \cite{Bellissard}.}. Matrix $\Delta$ as well
 as matrix $G$ can be decomposed into following blocs
\begin{equation}
\Delta=\left( \matrix{A&B\cr
                  B^t&D} \right),
\end{equation}
\begin{equation}
G=\left( \matrix{a&b\cr
                  b^t&d} \right),
\end{equation}
where the infinite-dimensional matrices $A,B,D,a,b,d$ are defined as follows. $A, B, D$ defined by the quadratic part of action eqs. (\ref{ext}, \ref{cross}, \ref{int}) (Fig. 5(a), 5(b), 5(c)). The matrices $d,b$ are the usual propagator matrices (Fig. 5(b), 5(c))  and $a$ is the inverse propagator between the points of sublattice, that is an expression inverse to $G(X)$. Now we can invert $D$ using the fact that matrices $\Delta$ and $G$ are inverse to each other. This implies a set of algebraic relations for their submatrices: 
$$
A a + B b^t=1,\,
A b + B d=0,\,
B^t a + D b^t=0,\,
B^t b + D d=1
$$
and, after straightforward transformations we obtain an expression for the "internal" propagator:
\begin{equation}
D^{-1}=d-b^t a b
\end{equation}
or, returning to previous notations,
\begin{equation}
\label{int-prop}
D^{-1}_{xy}=G_{xy}-\sum_{X,Y}G_{xX} {\bf G}^{-1}_{XY} G_{Yy} 
\end{equation}
\begin{picture}(10000,3000)(0,-2000)
\put(5000,-300){$=$}
\drawline\fermion[\E\REG](7000,0)[5000]
\put(13000,-300){$-$}
\drawline\fermion[\E\REG](16000,0)[5000]
\THICKLINES
\drawline\fermion[\E\REG](\pbackx,0)[5000]
\THINLINES
\put(\fermionfrontx,0){\circle{500}}
\drawline\fermion[\E\REG](\pbackx,0)[5000]
\put(\fermionfrontx,0){\circle{500}}
\end{picture}

Here and in what follows bold line denotes free inverse decimated propagator ${\bf G}^{-1}_{XY}$, while thin line corresponds to propagator of the original theory $G_{xy}$.
Using this representation of $D^{-1}_{xy}$ as an internal line, we can now begin to build the perturbation theory. We would like to stress out that using of $D^{-1}_{xy}$ eq. (\ref{int-prop}) enables us to extend this summation over $\it all$ the original lattice ${\cal L}$ rather then "decorated" subset ${\cal D}$. Indeed, one can see that this expression is equal to zero when at least one of the points $x, y$ in $D^{-1}_{xy}$ belongs to ${\cal L}^*$ (Fig. 6). Therefore, any additional contribution due to this extension is equal to zero as well.

\begin{picture}(40000,13000)(0,-7500)

\put(5000,2000){\circle{5000}}
\drawline\fermion[\SE\REG](5000,0)[3000]
\drawline\fermion[\SW\REG](5000,0)[3000]
\drawline\fermion[\S\REG](5000,0)[3000]
\put(5000,0){\circle*{5000}}
\put(5000,4000){\circle{500}}
\put(10000,1000){$-$}
\put(16000,2000){\circle{5000}}
\drawline\fermion[\SE\REG](16000,0)[3000]
\drawline\fermion[\SW\REG](16000,0)[3000]
\drawline\fermion[\S\REG](16000,0)[3000]
\put(16000,0){\circle*{5000}}
\put(18000,2000){\circle{500}}
\put(14000,2000){\circle{500}}
\put(16000,4000){\circle{500}}
\put(21000,1000){$= \;\; 0$}

\put(0,-4000){Fig. 6. An additional contribution from extended summation.}
\put(3000,-5500){ Small circles denote the points of the sublattice ${\cal L}^*$.}
\end{picture}

Calculation of an n-point function $H^{(n)}(X_1,..,X_n)$ in effective action eq. (\ref{acteff}) for the coarse-grained field $\phi_X$ will be as follows. 

 All connected diagrams with $n$ end points $X_1,..,X_n$, $V$ vertices and $I$ internal lines in real space are drawn with following components:

a) all vertices are situated at the points $x\in {\cal L}$ and to every vertex at point $x_i$ corresponds summation $\sum_{x_i}$;

b) $\dbar_{Xx}$ are assigned to external ends $X$ and

c) the internal lines $D^{-1}_{xy}$ are represented via eq.(\ref{int-prop}). This representation splits each diagram into $2^I$ subdiagrams and each of these subdiagrams should be calculated separately. This calculation includes summation over all the internal points $x_i$ on the fine grained lattice and over the internal sublattice points $Y$ [end points of the inverse decimated propagator ${\bf G}^{-1}$, see eq. (\ref{int-prop})]. Finally the "classical contribution" to coefficient function should be added.

For the sake of simplicity let us discuss calculations of the decimation diagrams on the concrete example. Namely, we will consider one of the contributions to four point function $H^{(4)}$ in $\phi^4$ model (Fig. 7).

With using of the propagator $D^{-1}_{xy}$, original diagram splits into five subdiagrams (Fig. 7(a), 7(b), 7(c), 7(d), 7(e)). Typical subdiagram here (for instance, subdiagram 7(d)) can be written as
\pagebreak 
$$
H^{(4)}_d (X_1,..,X_4)=
$$
\begin{equation}
\sum\limits_{\scriptstyle y_1,..,y_4,z \atop \scriptstyle Y_1,..Y_4} \dbar_{X_1\; y_1} G_{y_1\; Y_1} {\bf G}^{-1}_{Y_1\; Y_2} G_{Y_2\; z} \dbar_{X_2\; y_2} G_{y_2\; Y_3} {\bf G}^{-1}_{Y_3\; Y_4} G_{Y_4\; z} 
\dbar_{X_3\; y_3} G_{x_3\; z} \dbar_{X_4\; y_4} G_{x_4\; z}.
\label{fourp} 
\end{equation}
As usual, in practical calculations it is very convenient to employ the Fourier transformed functions at the intermediate steps. In this way we will operate with vertices, legs    
\begin{equation}
\dbar(k)=2 a^{d-2} \sum_\mu {\rm cos}(k_\mu a),
\end{equation}
propagators $G(k)$ and inverse decimated propagator ${\bf G}^{-1}(K)$ eq. (\ref{decprop}). However, it is not convenient to perform Fourier transform of expression eq. (\ref{fourp}) immediately, because it contains functions defined on different lattices and, therefore, obeying different transformation rules (cf. eqs. (\ref{finefourier1},\ref{finefourier2},\ref{coarsefourier1},\ref{coarsefourier2})).

 Instead, we can use the fact that this expression breaks into blocks where all internal points lie on the fine grained lattice ${\cal L}$ and sublattice points enter only as ends. These blocks are connected by ${\bf G}^{-1}(Y_i,Y_j)$. In terms of such a "${\cal L}$-connected" parts, diagram eq. (\ref{fourp}) has the form:  
$$
H^{(4)}_d (X_1,..,X_4)
$$
\begin{equation}
\label{blockdiagr}
=\sum\limits_{ \scriptstyle Y_1,..Y_4} F_1(X_1,\; Y_1) F_2(X_2,\; Y_2) F_3(X_3,\;X_4,\;Y_3,\; Y_4){\bf G}^{-1}_{Y_1\; Y_2}{\bf G}^{-1}_{Y_3\; Y_4}
\end{equation}
with blocks
\begin{equation}
F_1(X_1,\; Y_1)=\sum\limits_{y_1} \dbar_{X_1\; y_1} G_{y_1\; Y_1},
\end{equation}
\begin{equation}
F_2(X_2,\; Y_2)= \sum\limits_{y_2} \dbar_{X_2\; y_2} G_{y_2\; Y_3}
\end{equation}
and
\begin{equation}
F_3(X_3,\;X_4,\;Y_3,\; Y_4)=\sum\limits_{y_3,y_4,z} G_{Y_2\; z} G_{Y_4\; z} 
\dbar_{X_3\; y_3} G_{x_3\; z} \dbar_{X_4\; y_4} G_{x_4\; z}.
\end{equation}

Now we can proceed as follows. First we will Fourier transform each block in eq. (\ref{blockdiagr}) with respect to its internal points. This will result in diagram $G^{(4)}_d$ as function on sublattice, and we can apply to it the Fourier transformation rules for the sublattice eq. (\ref{coarsefourier1},\ref{coarsefourier2}). First step does not differ from the usual perturbative lattice calculations, and resulting functions read, for example, as 
\begin{equation}
F_1(X_1,\; Y_1)=\int_{-\pi/a}^{\pi/a} {\rm e}^{i(X_1-Y_1)k} \dbar(k)G(k).
\end{equation}

\begin{picture}(45000,46000)(0,-33000)
\drawline\fermion[\SE\REG](0,10000)[9000]
\drawline\photon[\NW\REG](\fermionfrontx,\fermionfronty)[2]
\put(\pbackx,\pbacky){\circle{500}}
\drawline\photon[\SE\REG](\fermionbackx,\fermionbacky)[2]
\put(\pbackx,\pbacky){\circle{500}}
\drawline\fermion[\NE\REG](0,\fermionbacky)[9000]
\global\Yone=\pmidy
\drawline\photon[\SW\REG](\fermionfrontx,\fermionfronty)[2]
\put(\pbackx,\pbacky){\circle{500}}
\drawline\photon[\NE\REG](\fermionbackx,\fermionbacky)[2]
\put(\pbackx,\pbacky){\circle{500}}

\global\advance \Yone by -300
\global\advance \fermionfrontx by 11000
\put(\fermionfrontx,\Yone){$\longrightarrow$}

\drawline\fermion[\SE\REG](20000,10000)[9000]
\drawline\photon[\NW\REG](\fermionfrontx,\fermionfronty)[2]
\put(\pbackx,\pbacky){\circle{500}}
\drawline\photon[\SE\REG](\fermionbackx,\fermionbacky)[2]
\put(\pbackx,\pbacky){\circle{500}}
\drawline\fermion[\NE\REG](20000,\fermionbacky)[9000]
\drawline\photon[\SW\REG](\fermionfrontx,\fermionfronty)[2]
\put(\pbackx,\pbacky){\circle{500}}
\drawline\photon[\NE\REG](\fermionbackx,\fermionbacky)[2]
\put(\pbackx,\pbacky){\circle{500}}
\put(22500,0){(a)}

\put(0,-7500){$-$} 

\put(2000,-7500){$\left ( \right.$}

\drawline\fermion[\SE\REG](5000,-4000)[1500]
\global\Ytwo=\fermionbacky
\drawline\photon[\NW\REG](\fermionfrontx,\fermionfronty)[2]
\put(\pbackx,\pbacky){\circle{500}}
\THICKLINES
\drawline\fermion[\SE\REG](\fermionbackx,\fermionbacky)[1500]
\THINLINES
\put(\pfrontx,\pfronty){\circle{500}}
\drawline\fermion[\SE\REG](\pbackx,\pbacky)[6000]
\put(\fermionfrontx,\fermionfronty){\circle{500}}
\drawline\photon[\SE\REG](\fermionbackx,\fermionbacky)[2]
\put(\pbackx,\pbacky){\circle{500}}
\drawline\fermion[\NE\REG](5000,\fermionbacky)[9000]
\drawline\photon[\SW\REG](\fermionfrontx,\fermionfronty)[2]
\put(\pbackx,\pbacky){\circle{500}}
\drawline\photon[\NE\REG](\fermionbackx,\fermionbacky)[2]
\put(\pbackx,\pbacky){\circle{500}}
\put(7500,-14000){(b)}

\global\advance \fermionbackx by 1000
\put(\fermionbackx,-7500){$+$ ... $\left. \right )$}

\put(17500,-7500){$+$}

\put(20000,-7500){$\left ( \right.$}

\drawline\fermion[\SE\REG](23000,-4000)[1500]
\drawline\photon[\NW\REG](\fermionfrontx,\fermionfronty)[2]
\put(\pbackx,\pbacky){\circle{500}}
\THICKLINES
\drawline\fermion[\SE\REG](\fermionbackx,\fermionbacky)[1500]
\THINLINES
\put(\pfrontx,\pfronty){\circle{500}}
\drawline\fermion[\SE\REG](\pbackx,\pbacky)[6000]
\put(\pfrontx,\pfronty){\circle{500}}
\drawline\photon[\SE\REG](\fermionbackx,\fermionbacky)[2]
\put(\pbackx,\pbacky){\circle{500}}
\drawline\fermion[\NE\REG](23000,\fermionbacky)[1500]
\drawline\photon[\SW\REG](\fermionfrontx,\fermionfronty)[2]
\put(\pbackx,\pbacky){\circle{500}}
\THICKLINES
\drawline\fermion[\NE\REG](\fermionbackx,\fermionbacky)[1500]
\THINLINES
\put(\pfrontx,\pfronty){\circle{500}}
\drawline\fermion[\NE\REG](\pbackx,\pbacky)[6000]
\put(\pfrontx,\pfronty){\circle{500}}
\drawline\photon[\NE\REG](\fermionbackx,\fermionbacky)[2]
\put(\pbackx,\pbacky){\circle{500}}
\put(25000,-14000){(c)}

\put(30000,-7500){$+$ ... $\left. \right )$}

\put(0,-21500){$-$}

\put(2000,-21500){$\left( \right. $}

\drawline\fermion[\SE\REG](5000,-18000)[1500]
\drawline\photon[\NW\REG](\fermionfrontx,\fermionfronty)[2]
\put(\pbackx,\pbacky){\circle{500}}
\THICKLINES
\drawline\fermion[\SE\REG](\fermionbackx,\fermionbacky)[1500]
\THINLINES
\put(\pfrontx,\pfronty){\circle{500}}
\drawline\fermion[\SE\REG](\pbackx,\pbacky)[3000]
\put(\pfrontx,\pfronty){\circle{500}}

\THICKLINES
\drawline\fermion[\SE\REG](\fermionbackx,\fermionbacky)[1500]
\THINLINES
\put(\pfrontx,\pfronty){\circle{500}}
\drawline\fermion[\SE\REG](\pbackx,\pbacky)[1500]
\put(\pfrontx,\pfronty){\circle{500}}

\drawline\photon[\SE\REG](\fermionbackx,\fermionbacky)[2]
\put(\pbackx,\pbacky){\circle{500}}
\drawline\fermion[\NE\REG](5000,\photonfronty)[1500]
\drawline\photon[\SW\REG](\fermionfrontx,\fermionfronty)[2]
\put(\pbackx,\pbacky){\circle{500}}
\THICKLINES
\drawline\fermion[\NE\REG](\fermionbackx,\fermionbacky)[1500]
\THINLINES
\put(\pfrontx,\pfronty){\circle{500}}
\drawline\fermion[\NE\REG](\pbackx,\pbacky)[6000]
\put(\pfrontx,\pfronty){\circle{500}}

\drawline\photon[\NE\REG](\fermionbackx,\fermionbacky)[2]
\put(\pbackx,\pbacky){\circle{500}}
\put(7500,-28000){(d)}

\global\advance \fermionbackx by 1000
\put(\fermionbackx,-21500){$+$ ... $\left. \right )$}

\put(18000,-21500){$+$}

\drawline\fermion[\SE\REG](23000,-18000)[1500]
\drawline\photon[\NW\REG](\fermionfrontx,\fermionfronty)[2]
\put(\pbackx,\pbacky){\circle{500}}
\THICKLINES
\drawline\fermion[\SE\REG](\fermionbackx,\fermionbacky)[1500]
\THINLINES
\put(\pfrontx,\pfronty){\circle{500}}
\drawline\fermion[\SE\REG](\pbackx,\pbacky)[3000]
\put(\pfrontx,\pfronty){\circle{500}}

\THICKLINES
\drawline\fermion[\SE\REG](\fermionbackx,\fermionbacky)[1500]
\THINLINES
\put(\pfrontx,\pfronty){\circle{500}}
\drawline\fermion[\SE\REG](\pbackx,\pbacky)[1500]
\put(\pfrontx,\pfronty){\circle{500}}

\drawline\photon[\SE\REG](\fermionbackx,\fermionbacky)[2]
\put(\pbackx,\pbacky){\circle{500}}
\drawline\fermion[\NE\REG](23000,\photonfronty)[1500]
\drawline\photon[\SW\REG](\fermionfrontx,\fermionfronty)[2]
\put(\pbackx,\pbacky){\circle{500}}
\THICKLINES
\drawline\fermion[\NE\REG](\fermionbackx,\fermionbacky)[1500]
\THINLINES
\put(\pfrontx,\pfronty){\circle{500}}
\drawline\fermion[\NE\REG](\pbackx,\pbacky)[3000]
\put(\pfrontx,\pfronty){\circle{500}}
\THICKLINES
\drawline\fermion[\NE\REG](\fermionbackx,\fermionbacky)[1500]
\THINLINES
\put(\pfrontx,\pfronty){\circle{500}}
\drawline\fermion[\NE\REG](\pbackx,\pbacky)[1500]
\put(\pfrontx,\pfronty){\circle{500}}

\drawline\photon[\NE\REG](\fermionbackx,\fermionbacky)[2]
\put(\pbackx,\pbacky){\circle{500}}
\put(25000,-28000){(e)}
\put(0,-31000){Fig. 7. Four point contribution to the $\phi^4$ effective action}

\end{picture}

This, however is not the case when the remaining Fourier transforms and summations over internal sublattice points are performed. Due to the difference between the Brillouin zones result will be sums rather then monomials. For instance, for the block $F_1$ we will obtain:
\begin{equation}
 F_1(K)= \sum_{all \; n_\mu=1}^\eta \left[\sum_\mu \frac {2}{\eta^{(d-2)}} {\rm cos}\left(\frac{K_\mu+2 \pi n_\mu}{\eta}\right)\right] \frac {1}{\eta^2  \sum_\mu 4{\rm sin}^2 (\frac{K_\mu+2 \pi n_\mu}{2\eta})} 
\end{equation}
Such an expressions we will call "decimated" and denote by $[| "expr" |]$:
\begin{equation}
[|f(K,L,...)|]\equiv \sum_{(n_{K\mu},n_{L\mu},..)=1}^\eta f(K+2 \pi n_K,L+2 \pi n_L,...)
\label{decimator}
\end{equation} 
Notice that the decimated function does not possess original translation invariance (with the period $2 \pi /a$) but instead turns out to be $2 \pi /(\eta a)$ periodic. 
Eventually, diagram $G^{(4)}_d$ takes the following form:
$$
H^{(4)}_d(K,P, Q)
$$
\begin{equation}
=[|F_1(K)|]{\bf G}^{-1}(K)[|F_1(P)|]{\bf G}^{-1}(P)[|F_3(K,P,Q,-K-P-Q)|].
\end{equation}
Another simple example, diagrammatic calculation of the $d=1$ free boson effective action, is given in Appendix B. 

Note that described algorithm is applicable not only to the tree level contributions. When the representation of the propagator eq (\ref{int-prop}) is used, some of the loop diagrams will contain ${\bf G}^{-1}$. If this is the case, one should calculate the corresponding decimated functions, and only then perform the loop integration. We will encounter such diagrams in the next section.

\section{$O(N)$ symmetric classical spins: weak coupling expansion}

In this section we apply the formalism developed in the previous section to the $O(N)$ symmetric nonlinear $\sigma$ model.  In $d$ dimensions, this model is described by action
\begin{equation}
A= -\frac {1}{2 g} \sum_x S^a\Box S^a,
\end{equation}
where $S^a$ ($a=1,..,N$) is $O(N)$ vector normalized on unity, $S^2=1$, $g$ is the coupling constant or temperature and $\Box$ is the lattice Laplacian. The partition function of this model is given by the path integral
\begin{equation}
Z= \int \prod_{x,a} dS^a_x \delta (S^2_x-1) \exp {\left[\frac {1}{2 g} \sum_x S^a\Box S^a\right]}.
\end{equation}
This is an example of a theory with constraints. To develop a perturbation theory for this model, it is convenient to re-express it in terms of the unconstrained fields \cite{Elitzur} $\pi^i$, $i=1,..,N-1$. For this purpose one can solve constraint for $S^N$, obtaining $S^a=(\sqrt{g} a^{(d-2)/2} \pi^i, \sqrt{1-g a^{(d-2)} \pi^2})$, and then the partition function in terms of "pions" $\pi^i$ will have the form:
$$
Z= \int \prod_{x,i} \frac {d\pi^i_x }{\sqrt {1-g a^{(d-2)} \pi^2_x}} \exp \left[\frac {1}{2} \sum_x a^{(d-2)} \left(\pi^i_x \Box \pi^i_x \right. \right.
$$
\begin{equation}
 \left. \left. +\frac {1}{g} \sum_x \sqrt {1-g a^{(d-2)} \pi^2_x} \Box \sqrt {1-g a^{(d-2)} \pi^2_x}\right)\right] .
\end{equation}

This last expression when used perturbatively gives rise to the infinite set of vertices both local (coming from the exponentiated and expanded measure) and via derivative couplings originating from the $S^N$ part of the action. For our present purpose it is sufficient to restrict diagrammatic to the order $g^2$ in coupling constant $g$. To this order, partition function takes the form:

$$
Z= \int \prod_{x,i} d \pi^i_x \; {\rm exp}  \left[ \frac {a^{(d-2)}}{2} \sum_x \left(\pi^i \Box \pi^i+g \left( \frac {a^{(d-2)}}{4} \pi^2 \Box \pi^2 + \pi^2 \right) \right. \right.
$$
\begin{equation}
\label{sigaction}
\left. \left.+g^2 \left( \frac {a^{2\;(d-2)}}{8} (\pi^2)^2 \Box \pi^2+\frac {a^{(d-2)}}{2} (\pi^2 )^2 \right)+...\right) \right]
\end{equation}

From this expression, the basic diagrams will be: massless propagator
$$
G(k)=\frac {a^2}{\sum_\mu 4 {\rm sin}^2 (ak_\mu/2)}
$$
and vertices (Fig. 8).

\begin{picture}(35000,32000)(-5000,-25000)

\drawline\fermion[\E\REG](0,5000)[6000]
\put(\pmidx,\pmidy){\circle*{500}}
\put(200,500){$g\;a^{(d-2)}/2$}
\put(200,-2500){(a)}
\drawline\fermion[\SE\REG](10000,10000)[4000]
\drawline\fermion[\NE\REG](\fermionbackx,\fermionbacky)[4000]
\drawline\gluon[\S\REG](\fermionfrontx,\fermionfronty)[3]
\drawline\fermion[\SE\REG](\gluonbackx,\gluonbacky)[4000]
\drawline\fermion[\SW\REG](\gluonbackx,\gluonbacky)[4000]
\put(11000,0){$(g a^{2\;(d-2)}/8)\, \Box$}
\put(11000,-2500){(b)}
\drawline\fermion[\SE\REG](2000,-5000)[4000]
\put(\pbackx,-6500){$x$}
\drawline\fermion[\NE\REG](\pbackx,\pbacky)[4000]
\drawline\scalar[\S\REG](\pfrontx,\pfronty)[2]
\drawline\fermion[\SE\REG](\pbackx,\pbacky)[4000]
\drawline\fermion[\SW\REG](\pfrontx,\pfronty)[4000]
\put(\pfrontx,-13000){$y$}
\put(2000,-16000){$\frac{g^2 a^{2\;(d-2)}}{4} \delta (x-y)$}
\put(3000,-18500){(c)}
\drawline\fermion[\SE\REG](13000,-5000)[4000]
\put(\pbackx,-6500){$x$}
\drawline\fermion[\NE\REG](\pbackx,\pbacky)[4000]
\drawline\scalar[\S\REG](\pfrontx,\pfronty)[2]
\drawline\gluon[\E\REG](\pmidx,\pmidy)[3]
\drawline\fermion[\SE\REG](\scalarbackx,\scalarbacky)[4000]
\drawline\fermion[\SW\REG](\pfrontx,\pfronty)[4000]
\put(\pfrontx,-13000){$y$}
\drawline\fermion[\SE\REG](\gluonbackx,\gluonbacky)[4000]
\drawline\fermion[\NE\REG](\gluonbackx,\gluonbacky)[4000]
\put(\gluonfrontx,-16000){$\frac{g^2 a^{3\;(d-2)}}{16} \delta (x-y) \Box$}
\put(15000,-18500){(d)}

\put(-3000,-21000){Fig. 8. First-order (a, b) and second-order (c,d) vertices of $\sigma$-model. The curly}
\put(0,-22500){line stands for a lattice Laplacian, the broken line - for a $\delta$ function.}

\end{picture}

Now we can perform decimation perturbatively. First of all, one can see that the "classical" part of the action comes from the measure and the cross term $\frac {1}{g} \sum_{Xx} \sqrt{1-g\;a^{(d-2)} \pi^2_X} \dbar_{Xx} \sqrt{1-g\;a^{(d-2)} \pi^2_x}$. Indeed, other classical terms are:
$$
\frac {1}{2} \left[ \sum_X \left( a^{(d-2)} (\pi_X \Box_{diag} \pi_X)+\frac {1}{g} \sqrt{1-g\;a^{(d-2)} \pi^2_X} \Box_{diag} \sqrt{1-g\;a^{(d-2)} \pi^2_X} \right) \right]=\frac {1}{g}.
$$ 
The same cancellation takes place, of course, also for the diagonal terms of the "internal" part of the action. 
At the same time the cross term expands as
\pagebreak
$$
\frac {1}{g} \sum_{Xx} \sqrt{1-g\;a^{(d-2)} \pi^2_X} \dbar_{Xx} \sqrt{1-g\;a^{(d-2)} \pi^2_x}=
\sum_{Xx} \left[ -\frac {a^{(d-2)}}{2} \pi^2_X \dbar_{Xx} -\frac {a^{(d-2)}}{2} \pi^2_x \dbar_{Xx} \right.
$$
$$
+\left. g \left(\frac{a^{2(d-2)}}{4} \pi^2_X \dbar_{Xx} \pi^2_x -\frac {a^{2(d-2)}}{8} (\pi^2_X)^2 \dbar_{Xx} -\frac {a^{2(d-2)}}{8} (\pi^2_x)^2 \dbar_{Xx} \right) +\dots \right] 
$$
\begin{equation}
=\sum_X \left[ -a^{(d-2)} \pi^2_X  -\frac {g\;a^{2(d-2)}}{4} (\pi^2_X)^2 \right]+\sum_x \left[ -a^{(d-2)} \pi^2_x  -\frac {g \;a^{2(d-2)}}{4} (\pi^2_x)^2 \right] 
\end{equation}
$$
 + \frac{g\; a^{2(d-2)}}{4} \sum_{Xx} \pi^2_X \dbar_{Xx} \pi^2_x +\dots
$$

\begin{picture}(35000,32000)(-5000,-18000)

\drawline\fermion[\E\REG](6000,10000)[4000]
\put(\pfrontx,8500){$X$}
\drawline\photon[\W\REG](\pfrontx,\pfronty)[3]
\put(\pbackx,8500){$x$}
\put(\pbackx,\pbacky){\circle{500}}
\put(4000,5700){$\frac {1}{a}\pi_X \dbar (X-x)$} 
\put(4700,3000){(a)}

\put(19500,12000){\circle{500}}
\put(20500,12000){\circle{500}}
\put(22000,12000){$X$}
\drawline\photon[\S\REG](20000,12000)[3]
\put(22000,\pbacky){$x$}
\drawline\fermion[\SE\REG](\photonbackx,\photonbacky)[4000]
\drawline\fermion[\SW\REG](\photonbackx,\photonbacky)[4000]
\put(\pbackx,4000){$\frac {g}{4 a^2} \pi^2_X\, \dbar (X-x)$}
\put(18000,2000){(b)}

\drawline\fermion[\SE\REG](3000,1000)[4000]
\put(\pbackx,-500){$x$}
\drawline\fermion[\NE\REG](\pbackx,\pbacky)[4000]
\drawline\scalar[\S\REG](\pfrontx,\pfronty)[2]
\put(4000,\pmidy){$x$}
\drawline\photon[\E\REG](\pmidx,\pmidy)[3]
\put(10500,\pbacky){$X$}
\drawline\fermion[\SE\REG](\scalarbackx,\scalarbacky)[4000]
\drawline\fermion[\SW\REG](\pfrontx,\pfronty)[4000]
\put(\pfrontx,-7000){$y$}
\global\advance \photonbackx 500
\global\advance \photonbacky 500
\put(\photonbackx,\photonbacky){\circle{500}}
\global\advance \photonbacky -1000
\put(\photonbackx,\photonbacky){\circle{500}}
\put(1500,-10000){$\frac{g^2}{16 a^2} \pi^2_X \delta (x-y) \dbar (X-x)$}
\put(5000,-12500){(c)}

\drawline\scalar[\S\REG](18000,\scalarfronty)[2]
\drawline\photon[\E\REG](\pmidx,\pmidy)[3]
\put(22500,\pbacky){$x$}
\put(\scalarfrontx,-1000){$X$}
\put(\pfrontx,-7000){$Y$}
\drawline\fermion[\SE\REG](\photonbackx,\photonbacky)[4000]
\drawline\fermion[\NE\REG](\photonbackx,\photonbacky)[4000]
\global\advance \scalarbackx 500
\put(\scalarbackx,\scalarbacky){\circle{500}}
\global\advance \scalarbackx -1000
\put(\scalarbackx,\scalarbacky){\circle{500}}
\global\advance \scalarfrontx 500
\put(\scalarfrontx,\scalarfronty){\circle{500}}
\global\advance \scalarfrontx -1000
\put(\scalarfrontx,\scalarfronty){\circle{500}}
\put(16000,-10000){$\frac{g^2}{16 a^2}\pi^2_X \pi^2_Y \delta (X-Y) \dbar (X-x)$}
\put(\scalarfrontx,-12500){(d)}

\put(3000,-14500){Fig. 9. The lowest order $\sigma$-model sources.}

\end{picture}

Terms belonging to the sublattice ${\cal L}^*$ contribute to the classical action while the terms lying on ${\cal D}$ complete the remaining off-diagonal internal part to the usual lattice action. 

The basic diagrammatic elements are: free massless propagator $G(x-y)$,  vertices, coinciding with the usual $\sigma$-model vertices (Fig. 8), sources: external leg $2\;a^{(d-2)} \dbar(X-x)$ (Fig. 9(a)) and cross interaction (Fig. 9(b), 9(c), 9(d)) and "classical" terms
\begin{equation}
\label{classical}
-A_4^{cl}= \sum_X \left[ -a^{(d-2)} \pi_X^2 -\frac {g\;a^{2(d-2)}}{4} (\pi_X^2)^2+ \frac{g\;a^{(d-2)}}{2} \pi_X^2 \right].
\end{equation}
Notice, that unlike local theories like $\phi^4$, in $\sigma$-model the "classical effective action" contains infinite series of such decoupled terms.
Performing the Fourier transform according the rules Sect. 2, we obtain the corresponding functions in momentum space: propagator
$$
G(k)=\frac {4}{a^2} {\rm sin}^2(\frac {a\; k}{2}),
$$
vertices (Fig. 8(a), 8(b), 8(c), 8(d))
$$
\frac {g\;a^{(d-2)}}{2} \delta^{ij},\,\, \frac {g\;a^{2(d-2)}}{8} \delta^{ij} \delta^{kl}\; 4\;{\rm sin}^2 (\frac {a\; k}{2}),\,\, \frac {g^2\;a^{2(d-2)}}{4} \delta^{ij} \delta^{kl},
$$
$$
\frac {g^2\;a^{3(d-2)}}{16} \delta^{ij} \delta^{kl} \delta^{mn}\;4\;{\rm sin}^2 (\frac {a\; k}{2}),
$$
and sources (Fig. 9(a), 9(b), 9(c), 9(d))
$$
\pi^i_K [|\dbar (K),\,\, \frac {g\;a^{(d-2)}}{4} \pi^i_K \pi^j_L \delta^{ij} \delta^{kl} [|\dbar (K+L),
$$
$$ \frac {g^2\;a^{2(d-2)}}{16} \pi^i_K \pi^j_L \delta^{ij} \delta^{kl} \delta^{mn} [|\dbar (K+L),  
$$
$$ 
 \frac {g^2\;a^{2(d-2)}}{16} \pi^i_K \pi^j_L \pi^k_P \pi^l_Q \delta^{ij} \delta^{kl} \delta^{mn} [|\dbar (K+L+P+Q),
$$
where 
$$
\dbar (k)=2\;a^{(d-2)} {\rm cos}(a k),
$$
and "half-decimation" $[| \dbar$ means that once the "{\cal L}-connected" part will be formed, it should be decimated. The classical part remains, of course, constant.

\section{Soluble model: $d=1$ Heisenberg chain} 

Let us consider first the simplest nontrivial example: $d=1$ $\sigma$ model and decimation with $\eta=2$. We start from the two point function. The tree level contribution is, obviously, that of the free theory (see Fig. 17 in Appendix B), in this dimension
\begin{equation}
H^{(0)}_2 (K) =- \left( \frac{2}{a}-[|\dbar^2(K) g(K)|]+[|\dbar(K) G(K)|]^2 {\bf G}^{-1}(K) \right).
\end{equation}
This contribution can be calculated analytically, and one can see that it reduces to
\begin{equation}
H^{(0)}_2 (K) = \;{\bf G}^{-1}(K)=\;\frac{4}{(\eta a)^2} {\rm sin}(\frac {\eta a K}{2})=4\; {\rm sin}(\frac K{2}).
\end{equation}
This corresponds to the fact that the free massless bosonic action in $d=1$ is perfect.

One loop contribution to propagator consists of three classes of diagrams: diagrams coming from the measure, bubble diagrams and the self energy part. 
Measure gives the contribution shown on Fig. 10, which reduces to 
\begin{equation}
H^{(1)}_m(K)=-\frac{g}{a}\; (1+\frac{1}{\eta a} {\rm cos}^2(\frac{\eta a K}{2})).
\end{equation}

\begin{picture}(25000,21000)(0,-5000)
\put(2000,12700){$H^{(1)}_m(K)=$}
\put(10000,13000){\circle{500}}
\put(11000,13000){\circle{500}}
\put(13000,12700){$+$}
\drawline\photon[\E\REG](16000,13000)[3]
\put(16000,13000){\circle{500}}
\drawline\fermion[\E\REG](\photonbackx,13000)[8000]
\put(\pmidx,13000){\circle*{500}}
\drawline\photon[\E\REG](\fermionbackx,13000)[3]
\put(\photonbackx,\photonbacky){\circle{500}}
\put(2000,7700){$-\,\, 2$}
\drawline\photon[\E\REG](5000,8000)[3]
\put(\photonfrontx,8000){\circle{500}}
\drawline\fermion[\E\REG](\photonbackx,8000)[4000]
\THICKLINES
\drawline\fermion[\E\REG](\fermionbackx,8000)[3000]
\THINLINES
\put(\fermionfrontx,8000){\circle{500}}
\drawline\fermion[\E\REG](\pbackx,8000)[6000]
\put(\pfrontx,8000){\circle{500}}
\put(\pmidx,\pmidy){\circle*{500}}
\drawline\photon[\E\REG](\fermionbackx,8000)[3]
\put(\photonbackx,8000){\circle{500}}
\put(0,2700){$+$}

\drawline\photon[\E\REG](5000,3000)[3]
\put(\photonfrontx,3000){\circle{500}}
\drawline\fermion[\E\REG](\photonbackx,3000)[4000]
\THICKLINES
\drawline\fermion[\E\REG](\fermionbackx,3000)[3000]
\THINLINES
\put(\fermionfrontx,3000){\circle{500}}
\drawline\fermion[\E\REG](\pbackx,3000)[6000]
\put(\pfrontx,3000){\circle{500}}
\put(\pmidx,3000){\circle*{500}}
\THICKLINES
\drawline\fermion[\E\REG](\fermionbackx,3000)[3000]
\THINLINES
\put(\fermionfrontx,3000){\circle{500}}
\drawline\fermion[\E\REG](\pbackx,3000)[4000]
\put(\pfrontx,3000){\circle{500}}
\drawline\photon[\E\REG](\fermionbackx,3000)[3]
\put(\photonbackx,3000){\circle{500}}

\put(0,-3000){Fig. 10. Contribution from path integral measure to one loop effective action.}
\end{picture}

Bubble diagrams are shown on Fig. 11. They give the contribution 
\begin{equation}
H^{(1)}_b(K)=-\frac {g\; (N-1)}{2 a} {\rm sin}^2(\frac{\eta a K}{2}).
\end{equation}
Notice here that although these integrals can seem to be IR divergent, in fact this is not the case: all the divergences cancel. The cancellation is due to the fact that in decimation only short distance effects are involved; long range "tails" remain unaffected.

\begin{picture}(40000,24000)(-8000,-17000)
\put(-8000,2500){$H^{(1)}_b(K)= $}

\put(-500,0){\circle{500}}
\put(500,0){\circle{500}}
\drawline\photon[\N\REG](0,200)[2]
\global\Yone \pbacky
\global\advance \Yone by 2000
\put(\pbackx,\Yone){\circle{4000}}
\put(0,-2000){a}

\put(4000,2500){$-$}

\put(8000,0){\circle{500}}
\put(9000,0){\circle{500}}
\drawline\photon[\N\REG](8500,200)[2]
\global\Xone \pbackx
\global\Yone \pbacky
\global\advance \Yone by 2000
\put(\pbackx,\Yone){\circle{4000}}
\global\advance \Xone by -1000
\global\advance \Yone by 1700
\put(\Xone,\Yone){\circle{500}}
\global\advance \Xone by 2000
\put(\Xone,\Yone){\circle{500}}
\put(8500,-2000){b}

\put(-9500,-7000){$+$}

\drawline\photon[\E\REG](-7000,-10000)[2]
\put(\pfrontx,\pfronty){\circle{500}}
\drawline\fermion[\E\REG](\pbackx,\pbacky)[6000]
\drawline\gluon[\N\REG](\pmidx,\pmidy)[2]
\global\Xone \pbackx
\global\Yone \pbacky
\global\advance \Yone by 2000
\put(\pbackx,\Yone){\circle{4000}}
\global\advance \Xone by -1000
\global\advance \Yone by 1700
\put(\Xone,\Yone){\circle{500}}
\global\advance \Xone by 2000
\put(\Xone,\Yone){\circle{500}}
\drawline\photon[\E\REG](\fermionbackx,\fermionbacky)[2]
\put(\pbackx,\pbacky){\circle{500}}
\put(-2000,-12000){c}

\put(4500,-7000){$- \,\, 2$}

\drawline\photon[\E\REG](8000,-10000)[2]
\put(\pfrontx,\pfronty){\circle{500}}
\drawline\fermion[\E\REG](\pbackx,\pbacky)[1000]
\THICKLINES
\drawline\fermion[\E\REG](\pbackx,\pbacky)[1000]
\THINLINES
\put(\pfrontx,\pfronty){\circle{500}}
\drawline\fermion[\E\REG](\pbackx,\pbacky)[4000]
\put(\pfrontx,\pfronty){\circle{500}}
\global\advance \pfrontx by 1000
\drawline\gluon[\N\REG](\pfrontx,\pfronty)[2]
\global\Xone \pbackx
\global\Yone \pbacky
\global\advance \Yone by 2000
\put(\pbackx,\Yone){\circle{4000}}
\global\advance \Xone by -1000
\global\advance \Yone by 1700
\put(\Xone,\Yone){\circle{500}}
\global\advance \Xone by 2000
\put(\Xone,\Yone){\circle{500}}
\drawline\photon[\E\REG](\fermionbackx,\fermionbacky)[2]
\put(\pbackx,\pbacky){\circle{500}}
\put(13000,-12000){d}

\put(19500,-7000){$+$}

\drawline\photon[\E\REG](22000,-10000)[2]
\put(\pfrontx,\pfronty){\circle{500}}
\drawline\fermion[\E\REG](\pbackx,\pbacky)[1000]
\THICKLINES
\drawline\fermion[\E\REG](\pbackx,\pbacky)[1000]
\THINLINES
\put(\pfrontx,\pfronty){\circle{500}}
\drawline\fermion[\E\REG](\pbackx,\pbacky)[2000]
\put(\pfrontx,\pfronty){\circle{500}}
\global\advance \pfrontx by 1000
\drawline\gluon[\N\REG](\pfrontx,\pfronty)[2]
\global\Xone \pbackx
\global\Yone \pbacky
\global\advance \Yone by 2000
\put(\pbackx,\Yone){\circle{4000}}
\global\advance \Xone by -1000
\global\advance \Yone by 1700
\put(\Xone,\Yone){\circle{500}}
\global\advance \Xone by 2000
\put(\Xone,\Yone){\circle{500}}
\THICKLINES
\drawline\fermion[\E\REG](\fermionbackx,\fermionbacky)[1000]
\THINLINES
\put(\pfrontx,\pfronty){\circle{500}}
\drawline\fermion[\E\REG](\pbackx,\pbacky)[1000]
\put(\pfrontx,\pfronty){\circle{500}}
\drawline\photon[\E\REG](\fermionbackx,\fermionbacky)[2]
\put(\pbackx,\pbacky){\circle{500}}
\put(27000,-12000){e}

\put(0,-14000){Fig. 11. One loop bubble diagrams.} 
\end{picture}

Last group, self energy diagrams, give the following contributions depicted on Fig. 12. This contribution reduces to
\begin{equation}
H^{(1)}_s(K)=\frac {g}{ a} {\rm cos}^2(\frac{\eta a K}{2}).
\end{equation}
One can see that again all the infrared divergences canceled.

\begin{picture}(45000,27000)(5000,-9000)
\put(0,15700){$ H^{(1)}_s(K)=$}

\put(0,12700){$=\quad \left( \right.$}

\drawline\photon[\E\REG](5000,13000)[2]
\put(\pfrontx,\pfronty){\circle{500}}
\drawline\fermion[\E\REG](\pbackx,\pbacky)[3000]
\drawline\fermion[\E\REG](\pbackx,\pbacky)[4500]
\global\advance \pmidy by -2000
\put(\pmidx,\pmidy){($a$)}
\drawloop\gluon[\NE 3](\pfrontx,\pfronty)
\drawline\fermion[\E\REG](\fermionbackx,\fermionbacky)[3000]
\drawline\photon[\E\REG](\pbackx,\pbacky)[2]
\put(\pbackx,\pbacky){\circle{500}}

\global\Xone=\pbackx
\global\advance \Xone by 2000
\put(\Xone,12700){$-$}
\global\advance \Xone by 2500

\drawline\photon[\E\REG](\Xone,13000)[2]
\put(\pfrontx,\pfronty){\circle{500}}
\drawline\fermion[\E\REG](\pbackx,\pbacky)[3000]
\drawline\fermion[\E\REG](\pbackx,\pbacky)[1500]
\drawloop\gluon[\NE 3](\pfrontx,\pfronty)
\THICKLINES
\drawline\fermion[\E\REG](\fermionbackx,\fermionbacky)[1500]
\THINLINES
\global\advance \pmidy by -2000
\put(\pmidx,\pmidy){($b$)}
\put(\pfrontx,\pfronty){\circle{500}}
\drawline\fermion[\E\REG](\fermionbackx,\fermionbacky)[1500]
\put(\pfrontx,\pfronty){\circle{500}}
\drawline\fermion[\E\REG](\fermionbackx,\fermionbacky)[3000]
\drawline\photon[\E\REG](\pbackx,\pbacky)[2]
\put(\pbackx,\pbacky){\circle{500}}

\global\advance \pbackx by 1500
\put(\pbackx,12700){$\left. \right)$}

\put(0,4700){$-\,\, 2\,\, \left( \right.$}

\drawline\photon[\E\REG](5000,5000)[2]
\put(\pfrontx,\pfronty){\circle{500}}
\drawline\fermion[\E\REG](\pbackx,\pbacky)[1000]
\THICKLINES
\drawline\fermion[\E\REG](\fermionbackx,\fermionbacky)[1000]
\THINLINES
\put(\pfrontx,\pfronty){\circle{500}}
\drawline\fermion[\E\REG](\fermionbackx,\fermionbacky)[1000]
\put(\pfrontx,\pfronty){\circle{500}}
\drawline\fermion[\E\REG](\pbackx,\pbacky)[4500]
\global\advance \pmidy by -2000
\put(\pmidx,\pmidy){($c$)}
\drawloop\gluon[\NE 3](\pfrontx,\pfronty)
\drawline\fermion[\E\REG](\fermionbackx,\fermionbacky)[3000]
\drawline\photon[\E\REG](\pbackx,\pbacky)[2]
\put(\pbackx,\pbacky){\circle{500}}

\global\Xone=\pbackx
\global\advance \Xone by 2000
\put(\Xone,4700){$-$}
\global\advance \Xone by 2500

\drawline\photon[\E\REG](\Xone,5000)[2]
\put(\pfrontx,\pfronty){\circle{500}}
\drawline\fermion[\E\REG](\pbackx,\pbacky)[1000]
\THICKLINES
\drawline\fermion[\E\REG](\fermionbackx,\fermionbacky)[1000]
\THINLINES
\put(\pfrontx,\pfronty){\circle{500}}
\drawline\fermion[\E\REG](\fermionbackx,\fermionbacky)[1000]
\put(\pfrontx,\pfronty){\circle{500}}
\drawline\fermion[\E\REG](\pbackx,\pbacky)[1500]
\drawloop\gluon[\NE 3](\pfrontx,\pfronty)
\THICKLINES
\drawline\fermion[\E\REG](\fermionbackx,\fermionbacky)[1500]
\THINLINES
\global\advance \pmidy by -2000
\put(\pmidx,\pmidy){($d$)}
\put(\pfrontx,\pfronty){\circle{500}}
\drawline\fermion[\E\REG](\fermionbackx,\fermionbacky)[1500]
\put(\pfrontx,\pfronty){\circle{500}}
\drawline\fermion[\E\REG](\fermionbackx,\fermionbacky)[3000]
\drawline\photon[\E\REG](\pbackx,\pbacky)[2]
\put(\pbackx,\pbacky){\circle{500}}

\global\advance \pbackx by 1500
\put(\pbackx,4700){$\left. \right)$}

\put(0,-3300){$+\quad \left( \right.$}

\drawline\photon[\E\REG](5000,-3000)[2]
\put(\pfrontx,\pfronty){\circle{500}}
\drawline\fermion[\E\REG](\pbackx,\pbacky)[1000]
\THICKLINES
\drawline\fermion[\E\REG](\fermionbackx,\fermionbacky)[1000]
\THINLINES
\put(\pfrontx,\pfronty){\circle{500}}
\drawline\fermion[\E\REG](\fermionbackx,\fermionbacky)[1000]
\put(\pfrontx,\pfronty){\circle{500}}
\drawline\fermion[\E\REG](\pbackx,\pbacky)[4500]
\global\advance \pmidy by -2000
\put(\pmidx,\pmidy){($e$)}
\drawloop\gluon[\NE 3](\pfrontx,\pfronty)
\drawline\fermion[\E\REG](\fermionbackx,\fermionbacky)[1000]
\THICKLINES
\drawline\fermion[\E\REG](\fermionbackx,\fermionbacky)[1000]
\THINLINES
\put(\pfrontx,\pfronty){\circle{500}}
\drawline\fermion[\E\REG](\fermionbackx,\fermionbacky)[1000]
\put(\pfrontx,\pfronty){\circle{500}}
\drawline\photon[\E\REG](\pbackx,\pbacky)[2]
\put(\pbackx,\pbacky){\circle{500}}

\global\Xone=\pbackx
\global\advance \Xone by 2000
\put(\Xone,-3300){$-$}
\global\advance \Xone by 2500

\drawline\photon[\E\REG](\Xone,-3000)[2]
\put(\pfrontx,\pfronty){\circle{500}}
\drawline\fermion[\E\REG](\pbackx,\pbacky)[1000]
\THICKLINES
\drawline\fermion[\E\REG](\fermionbackx,\fermionbacky)[1000]
\THINLINES
\put(\pfrontx,\pfronty){\circle{500}}
\drawline\fermion[\E\REG](\fermionbackx,\fermionbacky)[1000]
\put(\pfrontx,\pfronty){\circle{500}}
\drawline\fermion[\E\REG](\pbackx,\pbacky)[1500]
\drawloop\gluon[\NE 3](\pfrontx,\pfronty)
\THICKLINES
\drawline\fermion[\E\REG](\fermionbackx,\fermionbacky)[1500]
\THINLINES
\global\advance \pmidy by -2000
\put(\pmidx,\pmidy){($f$)}
\put(\pfrontx,\pfronty){\circle{500}}
\drawline\fermion[\E\REG](\fermionbackx,\fermionbacky)[1500]
\put(\pfrontx,\pfronty){\circle{500}}
\drawline\fermion[\E\REG](\fermionbackx,\fermionbacky)[1000]
\THICKLINES
\drawline\fermion[\E\REG](\fermionbackx,\fermionbacky)[1000]
\THINLINES
\put(\pfrontx,\pfronty){\circle{500}}
\drawline\fermion[\E\REG](\fermionbackx,\fermionbacky)[1000]
\put(\pfrontx,\pfronty){\circle{500}}
\drawline\photon[\E\REG](\pbackx,\pbacky)[2]
\put(\pbackx,\pbacky){\circle{500}}

\global\advance \pbackx by 1500
\put(\pbackx,-3300){$\left. \right)$}

\put(0,-7000){Fig. 12. Self energy-type contribution to one loop effective action}
\end{picture}

The total one loop quadratic part of an effective action with restored coupling constant therefore takes the following simple form:
\begin{equation}
A^{eff}(\pi_X)=\frac{g\;(N-1)}{16 a} \sum_{XY}\pi^i_X\Box(X-Y)\pi^i_Y-\frac{g}{2 a} \sum_{X}\pi^2_X,
\end{equation}
where $\Box(X)$ is understood as lattice operator. 

Our next step will be the four-point function. 

Let us consider next terms in an expansion of the effective action. They contain four fields and up to four derivatives [see eq. (\ref{sigaction})]. At the tree level, there are three contribution to these terms. First is a "classical" term $A_4^{cl}$ eq. (\ref{classical}). Two other contributions are given by diagrams Fig. 13, 14.

Again, using the representation eq. (\ref{int-prop}) splits every diagram into several.

\begin{picture}(40000,20000)(4000,-9000)

\drawline\fermion[\SE\REG](5000,5000)[3000]
\drawline\photon[\NW\REG](\fermionfrontx,\fermionfronty)[2]
\put(\pbackx,\pbacky){\circle{500}}
\drawline\fermion[\SW\REG](\fermionbackx,\fermionbacky)[3000]
\drawline\photon[\SW\REG](\fermionbackx,\fermionbacky)[2]
\put(\pbackx,\pbacky){\circle{500}}
\drawline\gluon[\E\REG](\fermionfrontx,\fermionfronty)[3]
\drawline\fermion[\SE\REG](\gluonbackx,\gluonbacky)[3000]
\drawline\photon[\SE\REG](\fermionbackx,\fermionbacky)[2]
\put(\pbackx,\pbacky){\circle{500}}
\drawline\fermion[\NE\REG](\gluonbackx,\gluonbacky)[3000]
\drawline\photon[\NE\REG](\fermionbackx,\fermionbacky)[2]
\put(\pbackx,\pbacky){\circle{500}}

\put(8000,-3500){(a)}

\drawline\fermion[\SE\REG](19000,7000)[3000]
\drawline\photon[\NW\REG](\fermionfrontx,\fermionfronty)[2]
\put(\pbackx,\pbacky){\circle{500}}
\drawline\fermion[\NE\REG](\fermionbackx,\fermionbacky)[3000]
\drawline\photon[\NE\REG](\fermionbackx,\fermionbacky)[2]
\put(\pbackx,\pbacky){\circle{500}}
\drawline\gluon[\S\REG](\fermionfrontx,\fermionfronty)[3]
\drawline\fermion[\SW\REG](\gluonbackx,\gluonbacky)[3000]
\drawline\photon[\SW\REG](\fermionbackx,\fermionbacky)[2]
\put(\pbackx,\pbacky){\circle{500}}
\drawline\fermion[\SE\REG](\gluonbackx,\gluonbacky)[3000]
\drawline\photon[\SE\REG](\fermionbackx,\fermionbacky)[2]
\put(\pbackx,\pbacky){\circle{500}}

\put(20000,-3500){(b)}

\drawline\fermion[\SE\REG](31000,7000)[3000]
\drawline\photon[\NW\REG](\fermionfrontx,\fermionfronty)[2]
\put(\pbackx,\pbacky){\circle{500}}

\drawline\gluon[\S\REG](\fermionbackx,\fermionbacky)[3]
\drawline\fermion[\SW\REG](\gluonbackx,\gluonbacky)[3000]
\drawline\photon[\SW\REG](\fermionbackx,\fermionbacky)[2]
\put(\pbackx,\pbacky){\circle{500}}
\drawline\fermion[\SE\REG](\gluonfrontx,\gluonfronty)[7000]
\drawline\photon[\SE\REG](\fermionbackx,\fermionbacky)[2]
\put(\pbackx,\pbacky){\circle{500}}
\drawline\fermion[\NE\REG](\gluonbackx,\gluonbacky)[7000]
\drawline\photon[\NE\REG](\fermionbackx,\fermionbacky)[2]
\put(\pbackx,\pbacky){\circle{500}}

\put(33000,-3500){(b)}

\put(5000,-6000){Fig. 13. Fourth-order contributions including internal vertex Fig. 8(b).}

\end{picture}

\begin{picture}(40000,17000)(-1000,-12000)

\put(-500,500){\circle{500}}
\put(-500,-500){\circle{500}}
\put(-2500,1000){$K,i$}
\put(-2500,-1500){$L,j$}
\drawline\photon[\E\REG](0,0)[3]
\put(500,1000){$K+L$}
\drawline\fermion[\SE\REG](\photonbackx,\photonbacky)[4000]
\drawline\photon[\SE\REG](\pbackx,\pbacky)[3]
\put(\pbackx,\pbacky){\circle{500}}
\put(8000,5500){$M,k$}
\drawline\fermion[\NE\REG](\fermionfrontx,\fermionfronty)[4000]
\drawline\photon[\NE\REG](\pbackx,\pbacky)[3]
\put(\pbackx,\pbacky){\circle{500}}
\put(4000,-8000){(a)}

\put(16500,4500){\circle{500}}
\put(17500,4500){\circle{500}}
\put(14000,4500){$K,i$}
\put(18000,4500){$M,k$}
\drawline\photon[\S\REG](17000,4000)[3]
\put(18000,\pmidy){$K+M$}
\drawline\fermion[\SW\REG](\photonbackx,\photonbacky)[4000]
\drawline\photon[\SW\REG](\pbackx,\pbacky)[3]
\put(\pbackx,\pbacky){\circle{500}}
\put(11500,-5500){$L,j$}
\drawline\fermion[\SE\REG](\fermionfrontx,\fermionfronty)[4000]
\drawline\photon[\SE\REG](\pbackx,\pbacky)[3]
\put(\pbackx,\pbacky){\circle{500}}

\put(17000,-8000){(b)}

\put(30500,4500){\circle{500}}
\put(31500,4500){\circle{500}}
\put(28000,4500){$L,j$}
\put(32000,4500){$M,k$}
\drawline\photon[\S\REG](31000,4000)[3]
\put(32000,\pmidy){$L+M$}
\drawline\fermion[\SW\REG](\photonbackx,\photonbacky)[4000]
\drawline\photon[\SW\REG](\pbackx,\pbacky)[3]
\put(\pbackx,\pbacky){\circle{500}}
\put(25500,-5500){$K,i$}
\drawline\fermion[\SE\REG](\fermionfrontx,\fermionfronty)[4000]
\drawline\photon[\SE\REG](\pbackx,\pbacky)[3]
\put(\pbackx,\pbacky){\circle{500}}

\put(31000,-8000){(c)}

\put(0,-11000){Fig. 14. The fourth-order contributions coming from the source Fig. 9(b).} 
\end{picture}
 
We calculate the coefficient function ${\bf H}^{(4)}$ as follows: an effective action as decimated one is given in momentum space by diagrams (Fig. 13, 14) plus classical contribution:

$$
-A^{(4)}_{eff}[\pi_X]=\int\limits_{K,L,M} (\pi^i_K \pi^i_L)(\pi^i_M \pi^i_{-K-L-M}) \frac {1}{4!} H^{(4)}_{ijkl}(K,L,M)
$$
\begin{equation}
\label{fourdec}
=\int\limits_{K,L,M} (\pi^i_K \pi^i_L)(\pi^i_M \pi^i_{-K-L-M}) \left[ -\frac {g}{4a}+({\rm diagram\; Fig. 13})+({\rm diagram\; Fig. 14}) \right],
\end{equation}
where
$$
({\rm diagram\; Fig. 13})=-(g/4!)\; \left[ \left( [|\dbar (K) \dbar (L) \dbar (M) \right. \right.
$$
\begin{equation}
\times \dbar (-K-L-M) G(K) G(L) G(M) G(-K-L-M) G^{-1}(K+L)|] 
\end{equation}
$$
\left. \left. +({\rm terms\; due\; to\; (\ref{int-prop})}) \right) +({\rm permutations}) \right],
$$
\begin{equation}
({\rm diagram\; Fig. 14})=(2 g/4!)\; \left[ \left( [|\dbar (K) \dbar (L) \dbar (K+L) G(K) G(L)|] \right. \right.
\end{equation}
$$
\left. \left. +({\rm terms\; due\; to\; (\ref{int-prop})}) \right) +({\rm permutations}) \right].
$$
On the other hand, this term in an effective action on the coarse grained lattice in one dimension could have the only form:
\begin{equation}
\label{foureff}
-A^{(4)}_{eff}[\pi_X]=- \int\limits_{K,L,M} \frac {1}{4!} {\bf H}^{(4)}_{ijkl}(K,L,M) \pi^i_K \pi^j_L \pi^k_M \pi^l_{-K-L-M} 
\end{equation}
with
\begin{equation}
{\bf H}^{(4)}_{ijkl}(K,L,M)=  {\bf H}^1_{ijkl} (K,L,M)+ c\; {\bf H}^2_{ijkl} (K,L,M). 
\end{equation}
Here the first term,
$$
{\bf H}^1_{ijkl} (K,L,M)=(g/4!) \left( \delta_{ij} \delta_{kl} (-4\; {\rm sin}^2((K+L)/2))+ ({\rm permutations})\right),
$$
was determined in previous subsections (it comes from expansion of the covariant action in terms of pions), and the second term is
$$
{\bf H}^2_{ijkl} (K,L,M)= -(g/4!)\; \left( \delta_{ij} \delta_{kl} ( 4\; {\rm sin}^2(K/2))( 4\; {\rm sin}^2(M/2)) \right.
$$
$$
\left. + ({\rm permutations})\right).
$$
To find the remaining coefficient $c$, we equate the coefficient functions ($\equiv$ fourth functional derivatives in fields $\pi^i_K$) in expressions eqs. (\ref{fourdec},\ref{foureff}) at some definite momentum configuration, for example, for $K=-L=M=P$ (the case of "back-to-back scattering") and for definite flavour indices ($i=j=k=l=1$, for instance):
\begin{equation}
H^{(4)}_{1111}(P,-P,P)={\bf H}^{(4)}_{1111}(P,-P,P).
\end{equation}
Then the coefficient $c$ is determined by the linear equation
\pagebreak
$$
\left( {d^4 H^{(4)}_{1111}(P,-P,P)}\over {d P^4} \right)_{P=0}=\left( {d^4 {\bf }H^1_{1111}(P,-P,P)}\over {d P^4} \right)_{P=0}
$$
\begin{equation}
\label{coeff}
+c\; \left( {d^4 H^2_{1111}(P,-P,P)}\over {d P^4} \right)_{P=0}.
\end{equation}
For this configuration: 
\begin{equation}
H^{(4)}_{1111}(P,-P,P)=g \left( -12 - 12 {\rm cos}^4(P/2)+ 2\; (5+6\; {\rm cos}(P)+{\rm cos}(2\; P)) \right),
\end{equation}
\begin{equation}
{\bf H}^{(4)}_{1111}(P,-P,P)=g \left( -4\; {\rm sin}^2 (P)-384\; c\; {\rm sin}^4(P/2)\right).
\end{equation}
Equation eq. (\ref{coeff}) gives then:
\begin{equation}
c=1/32.
\end{equation}

Notice, that at the first sight the decimated expression, eq. (\ref{fourdec}), should behave as $1/P^4$ (one can check that using the representation (\ref{int-prop}) cancels leading term $1/P^6$ even without decimation). However, due to decimation procedure all the terms up to constant cancel, and in fact $H^{(4)}_{ijkl}(K,L,M)$ begins from the second order terms. Moreover, equality of the $P^2$ terms in $H^{(4)}_{1111}(P,-P,P)$ and ${\bf H}^{(4)}_{1111}(P,-P,P)$ can serve as some additional consistency check.

The one loop contributions to the four derivative terms are shown in Figs. 18, 19 and 20 (Appendix D). For the sake of simplicity, we restrict our consideration to the leading $1/N$ contribution. These diagrams are calculated analogously to the one loop two derivatives terms, and give the following contribution to the effective action:
\begin{equation}
\frac {g^2 \; (N-1)}{2} \sum_X ((\pi^2)^2-\frac{1}{8} \pi^2 \Box \pi^2). 
\end{equation}
Again, all the potential IR divergences cancel due to the decimation procedure.

To calculate the four point function, it is more convenient to employ somewhat different method. Namely, instead of direct calculation of the effective (decimated) action, we will use a matching approach. The clue to this approach is that decimation does not change field variables, so that their correlators between points of sublattice ${\cal L}^*$, calculated in both original (fine grained) and effective (coarse grained) models should coincide. We will therefore calculate correlators of fields in points $X_1,X_2,..,X_n \in {\cal L}^*$ starting from different scales, $a$ and $A$ in terms of the lattice spacing, and after this we will require matching conditions between these functions to be fulfilled. For practical calculations this means that we should compare the amplitudes under considerations for two cases: decimated with parameter $\eta=A/a$ for the original lattice model and the usual lattice amplitudes calculated from the "phenomenological" Lagrangian including irrelevant operators with as yet free coefficients:
\begin{equation}
G^{(n)}(\eta X_1,\eta X_2,..,\eta X_n)={\bf G}^{(n)}(X_1,X_2,..,X_n).
\end{equation}
In momentum space this condition takes the form
\begin{equation}
[|G^{(n)}(p,k,..)|]|_{(p=P, k=K,...)}={\bf G}^{(n)}(P,K,..).
\end{equation}
 In fact, the usual approach when the lattice quantities are compared to the continual ones is nothing but the matching for particular value $A \rightarrow \infty$, although its physical meaning is not as transparent as for the decimation RG.
Strictly speaking, any amplitude is given as a power series in momenta, so that we should truncate this series at some point and consider truncated effective Lagrangian with finite number of irrelevant operators rather than exact one. Matching conditions will then fix these coefficients. Here we will restrict our consideration to the four derivative terms in an effective action. In $d=1$ $\sigma$-model, there could be only one such term, $(S_X \Box S_Y)^2$. Thus, our "phenomenological Lagrangian" has the form:
\begin{equation}
L=   L_0   +\frac {c}{g} \; (S_X \Box (X-Y) S_Y)^2,
\end{equation}
where $L_0$ is the usual $\sigma$-model Lagrangian.

Because there is only one arbitrary coefficient in a decimated Lagrangian, we need the four point function calculated for one particular configuration of external momenta, for example $[|G^{(4)}_{ijkl}(P,P,-P)|]$ . By four point function we mean correlator of unconstrained fields, $\pi^i$, rather than constrained fields $S^a$. Then the second configuration (back-to-back scattering, $[|G^{(4)}_{ijkl}(P,-P,P)|]$, for instance) can be used as a consistency check. 

On the tree level, this amplitude as calculated from the scale $a$ is given by three contributions:
$$
[|G^{(4)}_{0\;ijkl}(K,L,M)|]=\delta_{ij} \delta_{kl} [|G^{(4)}_1 (K,L,M)|]+\delta_{ik} \delta_{jl} [|G^{(4)}_2 (K,L,M)|]
$$
$$
+\delta_{il} \delta_{jk} [|G^{(4)}_3(K,L,M)|]
$$
To perform matching, it is sufficient to consider the term $\delta_{ij} \delta_{kl} [|G^{(4)}_1 (p)|]$ only. This term is given by diagram Fig. 15 and for configuration $K=L=-M=P$ has the following expansion:
\begin{equation}
\label{aexpan}
[|G^{(4)}_1(P)|]=-\frac {4}{P^6}-\frac {1}{4\; P^4}+...
\end{equation}

\begin{picture}(25000,15000)(-10000,-5000)
\put(-7000,3300){$\delta_{ij} \delta_{kl} [|G^{(4)}_1(K,L,M)|]=$}
\put(9500,7000){\circle{500}}
\put(9500,700){\circle{500}}
\put(7000,7000){$K,\; i$}
\put(7000,700){$L,\; j$}
\drawline\fermion[\SE\REG](10000,7000)[4000]
\drawline\fermion[\NE\REG](10000,1000)[4000]
\drawline\gluon[\E\REG](13000,4000)[3]
\drawline\fermion[\NE\REG](16000,4000)[4000]
\drawline\fermion[\SE\REG](16000,4000)[4000]
\put(19000,7000){\circle{500}}
\put(19000,700){\circle{500}}
\put(20000,7000){$M,\; k$}
\put(20000,700){$l$}
\put(-3000,-2000){Fig. 15. Fine grained four point function}   
\end{picture}

The same correlator but calculated as a sublattice quantity, has the following form (Fig. 16):

\begin{picture}(35000,16000)(0,-5000)
\put(3000,9000){$\delta_{ij} \delta_{kl} {\bf G}^{(4)}(K,L,M)=$}
\put(1000,4000){$=$}
\put(2000,7000){$K,\; i$}
\put(2000,700){$L,\; j$}
\THICKLINES
\drawline\fermion[\SE\REG](5000,7000)[4000]
\drawline\fermion[\NE\REG](5000,1000)[4000]
\THINLINES
\drawline\gluon[\E\REG](8000,4000)[3]
\THICKLINES
\drawline\fermion[\NE\REG](11000,4000)[4000]
\drawline\fermion[\SE\REG](11000,4000)[4000]
\THINLINES
\put(15000,7000){$M,\; k$}
\put(15000,700){$l$}
\put(16000,3700){$+\, 2\; c$}
\put(19000,3700){$\left( \right.$}
\THICKLINES
\drawline\fermion[\SE\REG](21000,7000)[4000]
\global\advance \pmidy by -400
\put(\pmidx,\pmidy){$\|$}
\drawline\fermion[\SE\REG](\fermionbackx,\fermionbacky)[4000]
\global\advance \pmidy by -400
\put(\pmidx,\pmidy){$\|$}
\drawline\fermion[\NE\REG](21000,1000)[8000]
\put(28000,3700){$+$ permutations $\left. \right)$}
\THINLINES
\put(19000,7000){$K,\; i$}
\put(19000,700){$L,\; j$}
\put(27500,7000){$M,\; k$}
\put(27500,700){$-K-L-M,\; l$}
\put(0,-2000){Fig. 16. Four point function in effective theory. Strokes correspond to the Laplacian.} 

\end{picture}

$$
{\bf G}^{(4)}(K,L,M)=-{\bf G}(K) {\bf G}(L) {\bf G}^{-1}(K+L) {\bf G}(M) {\bf G}(-K-L-M)
$$
$$
 -2\; c\; \left[ {\bf G}(K) {\bf G}(M)+{\bf G}(K) {\bf G}(-K-L-M) \right.
$$
\begin{equation}
\left. +{\bf G}(L) {\bf G}(M)+{\bf G}(L) {\bf G}(-K-L-M) \right],
\end{equation}
and expands at $K=L=-M=P$ as
\begin{equation}
\label{Aexpan}
{\bf G}^{(4)}(P)=-\frac {4}{P^6}-\frac {8\; c}{P^4}+...
\end{equation}
Comparing these two expressions, eqs. (\ref{aexpan},\ref{Aexpan}), one can find $c=1/32$. Thus, the only tree level four-derivative term in the decimated effective action in $d=1$ is 
$$
\frac {g}{32 a}\sum_X (\pi_X \Box \pi_X)^2.
$$

In fact, in the case $\eta=2$ the problem becomes especially simple. In this case the partition function has the following form:
$$
Z= \int \prod_{x,i} d \pi^i_x \; {\rm exp}  \left[ \frac {1}{a} \left( -\sum_X \pi^2_X -\sum_x \pi^2_x +\sum_{X,x} \pi^i_X \dbar_{Xx} \pi^i_x \right) \right.
$$
$$\left. +g \left( \frac {1}{4a^2} \left( -\sum_X (\pi^2_X)^2 -\sum_x (\pi^2_x)^2 +\sum_{X,x} \pi^2_X \dbar_{Xx} \pi^2_x \right) + \frac {1}{2a} \sum_X \pi^2_X+\frac {1}{2a} \sum_x \pi^2_x \right) \right.
$$
$$
+g^2 \left( \frac {1}{8a^3} \left( -\sum_X (\pi^2_X)^3 -\sum_x (\pi^2_x)^3 +\frac {1}{2} \sum_{X,x} (\pi^2_X)^2 \dbar_{Xx} \pi^2_x +\frac {1}{2} \sum_{X,x} \pi^2_X \dbar_{Xx} (\pi^2_x)^2 \right)+ \right.
$$
\begin{equation}
\label{1daction}
 \left. \left. \frac {1}{4a^2} \sum_X (\pi^2_X )^2 +\frac {1}{4a^2} \sum_x (\pi^2_x )^2 \right)+... \right] .
\end{equation}
Therefore, for $\eta=2$ an internal line reduces to the contact term, $\Box_{xy}=-2 \delta_{xy}$ for $(x,y) \in {\cal D}$. This means that both an internal propagator and all the internal vertices are local, $G(k)=a^2$ ets (the problem becomes classical). The only remaining non-local terms are sources. Moreover, in this case ${\cal D}$ is a lattice itself and thus there is no need to use the representation eq. (\ref{int-prop}) (it becomes trivial as the only non zero contribution comes from the basic diagram without any replacement). The loops shrink into the points and therefore there is no loop integrals. Due to all these simplifications, all the calculations can be done immediately in the real space. 

We can compare our perturbative RG results with the perturbative effective action obtained from the partition function eq. (\ref{1daction}). One can see that these expressions coincide.

Effective action obtained here differs from the exact effective action for the $d=1$ $\sigma$ model \cite{Rosenstein}. This difference is not surprising, however. Indeed, an exact decimation takes into account both perturbative and instanton-like nonperturbative configurations while here we restrict our consideration to the perturbative contribution only.   

\section{Two dimensional $\sigma$-model}

The real space RG discussed in previous sections is not restricted to the $d=1$ models only, but is immediately generalized to the higher dimensional theories. In this case calculations become much more cumbersome but otherwise all the technique remains unchanged. Here we would like to consider one of the most popular two dimensional theories - $d=2$ nonlinear $O(N)$ $\sigma$-model.

Diagrammatics of the model under consideration coincides with one dimensional case. Therefore, we can immediately use diagrams described above and the formalism from Section 3 to calculate an effective action. The only differences are disappearance of factors $a^{(d-2)}$ from the action and, of course, two dimensional sums in place of one dimensional. In $d=2$, however, action is not perfect anymore and thus an exact analytical results not always can be obtained. Instead the numerical methods should be employed for this theory. Moreover, our discussion here will be restricted to the simplest case of decimation with the parameter $\eta =2$ on the tree level. Even such a simple transformation, however, can be of a use when the lattice calculations are concerned. 

Decimation diagrammatics once again consists of the internal propagator
\begin{equation}
G(k)=\frac{a^2}{\sum_{\mu} 4 {\rm sin}^2 \frac {a k_\mu}{2}},
\end{equation}
external legs
\begin{equation}
\dbar (k) =2 \sum_{\mu} {\rm cos} (a k_\mu),
\end{equation}
inverse free decimated propagator ${\bf G}^{-1}(K)$ and the vertices (Fig. 8, Fig. 9).

The diagrams similar to one-dimensional (see previous Section), or equivalently matching conditions, determine form of the coefficient function for the effective action's quadratic part ${\bf H}^{(2)}(K)$ for $\eta=2$ as
$$
{\bf H}^{(2)}(K)=-{\bf G}^{-1}(K),
$$
where
$$
{\bf G}^{-1}(K)=\left(\frac {1}{8\; (2+{\rm cos}(K_1/2)-{\rm cos}(K_2/2))}+\frac {1}{8\; (2-{\rm cos}(K_1/2)+{\rm cos}(K_2/2))} \right.
$$
\begin{equation}
\left.+\frac {1}{8\; (2+\sum_\mu {\rm cos}(K_\mu/2))}+\frac {1}{16\; \sum_\mu {\rm sin}^2 (K_\mu/4)}\right)^{-1}
\end{equation}
is an inverse decimated propagator.
Up to the order $O(K^4)$ it can be approximated as
\begin{equation}
{\bf H}^{(2)}(K)\approx -\sum_\mu 4 {\rm sin}^2 (K_\mu/2) -\frac {1}{16} \sum_\mu (4 {\rm sin}^2 (K_\mu/2))^2 +\frac {5}{32} (\sum_\mu 4 {\rm sin}^2 (K_\mu/2))^2.
\end{equation}
On the other hand, this term should have the form (see Appendix C):
$$
{\bf H}^{(2)}(K)=-\sum_\mu 4 {\rm sin}^2 (K_\mu/2)+(-\frac {1}{12}+2\; c_6) \sum_\mu (4 {\rm sin}^2 (K_\mu/2))^2+
$$
\begin{equation}
2\; c_5\; (\sum_\mu 4 {\rm sin}^2 (K_\mu/2))^2.
\end{equation}
Comparing these two expression, we find:
\begin{equation}
c_5=5/32; \quad c_6=1/96.
\end{equation}

Next, quartic terms in an effective action's expansion are calculated similarly to one dimensional case. Again, there are three contributions to these terms: classical term 
$$
-(g/8) \sum_X (\pi_X^2)^2
$$
and two different diagrams (Fig. 13, Fig. 14). An effective theory, however, allows now for three different four derivative terms (see Appendix C) contrary to the previous example where such term was unique. An effective action in momentum space has the form:
\begin{equation}
-A^{(4)}_{eff}[\pi_X]=-\frac{1}{2 \pi} \int\limits_{K,L,M} \frac {1}{4!} {\bf H}^{(4)}_{ijkl}(K,L,M) \pi^i_K \pi^j_L \pi^k_M \pi^l_{-K-L-M} 
\end{equation} 
with

$$
{\bf H}^{(4)}_{ijkl}(K,L,M)=  {\bf H}^1_{ijkl} (K,L,M)+ c_7\; {\bf H}^7_{ijkl} (K,L,M)+c_8\; {\bf H}^8_{ijkl} (K,L,M)
$$
\begin{equation}
+c_9\; {\bf H}^9_{ijkl} (K,L,M). 
\end{equation}
Once again the first term,
$$
{\bf H}^1_{ijkl} (K,L,M)=(g/4!) \left( \delta_{ij} \delta_{kl} (-{\bf G}^{-1}(K+L))+ ({\rm permutations})\right),
$$
comes from expansion of the quadratic part of the covariant action in terms of pions. Next terms are:

$$
{\bf H}^7_{ijkl} (K,L,M)= -(g/4!)\; \left( \delta_{ij} \delta_{kl} ( \sum_\mu 4\; {\rm sin}^2(K_\mu /2))(\sum_\mu 4\; {\rm sin}^2(M_\mu /2)) \right.
$$
$$
\left. + ({\rm permutations})\right),
$$
$$
{\bf H}^8_{ijkl} (K,L,M)= -(g/4!)\; \left( \delta_{ij} \delta_{kl} ( \sum_\mu 16\; {\rm sin}^2(K_\mu /2)\; {\rm sin}^2(M_\mu /2)) \right.
$$
$$
\left. + ({\rm permutations})\right),
$$
and 
$$
{\bf H}^9_{ijkl} (K,L,M)= -(g/4!)\; \left( \delta_{ij} \delta_{kl} ( \sum_{\mu ,\nu }{\rm sin}(K_\mu)\; {\rm sin}(L_\nu)\; {\rm sin}(M_\mu)\;  \right.
$$
$$
\left. \times {\rm sin}((-K-L-M)_\nu))) + ({\rm permutations})\right).
$$
To fully determine this quartic terms, we need now to fix three coefficients ( $c_7$, $c_8$ and $c_9$). This can be done calculating the coefficient functions $H^{(4)}$, ${\bf H}^{(4)}$ for three linearly independent momenta configurations. \footnote{Notice that technically decimation is nothing but summation over functions with an arguments shifted on $a\pi n$. As a consequence, decimated expressions including functions like $\dbar (k)$ can be sensitive to the sign of the argument even if the original expressions are not. Therefore, when calculating the decimated diagrams, one should take into account all different momenta permutations.} This will give us a system of three linear equations similar to eq. (\ref{coeff}) for the coefficients. It is convenient to chose the following configurations: 
$$
K=-L=M=(p,\; p),
$$
$$
K=(0,\; p),\; L=(p,\; -p),\; M=(p,\; 0),
$$
and 
$$
K=L=M=(p,\; 0).
$$
Then the coefficients $c_7,\; c_8,\; c_9$ will obey the equations:
$$
32/3 + 96\; c_7 + 48\; c_8 + 96\; c_9=19/6,
$$
\begin{equation}
245/48 + 78\; c_7 + 34\; c_8 + 16\; c_9=185/48,
\end{equation}
$$
17/2 + 120\; c_7 + 120\; c_8 - 72\; c_9=113/8
$$
with the solution:
\begin{equation}
c_7=c_8=0,\, c_9=-5/64
\end{equation}
The vanishing of $c_7$ and $c_8$ is rather surprising. We do not see any obvious reason for this. 

On the tree level, one can still employ matching method to calculate next coefficients in expansion of an effective action:
\begin{equation}
G^{(4)}_{ijkl}(\eta X,\eta Y,\eta Z)={\bf G}^{(4)}_{ijkl}( X,Y,Z).
\end{equation}
 Four point function as calculated in the original theory is given by the same diagram as in $d=1$ case. An effective theory, however, allows now for three different four derivative terms and thus there will be four contribution to the effective correlator ${\bf G}^{(4)}_{ijkl}$ instead of two (Fig. 16). Thus to determine all the coefficients, we need to impose matching conditions for at least three independent momenta configurations. Numerical calculations gave the following result (a momenta configurations have been chosen the same as before):

 For $P=-Q=R=(p,\; p)$;
\begin{equation}
[|G^{(4)}( P, Q, R)|]= -\frac {1/2}{p^6}-\frac {15/32}{p^4}+...,
\end{equation}
\begin{equation}
{\bf G}^{(4)}( P,Q,R)= -\frac {1/2}{p^6}+\frac {6\; c_7 + 3\; c_8 + 6\; c_9}{p^4}+...
\end{equation}

For $P=(0,p)$, $Q=(p,-p)$, $R=(p,0)$;
\begin{equation}
[|G^{(4)}(P, Q, R)|]= -\frac {1}{p^6}-\frac {117/128}{p^4}+...,
\end{equation}
\begin{equation}
{\bf G}^{(4)}( P,Q,R)= -\frac {1}{p^6}+\frac {-97/128 + (39/4)\; c_7 + (17/4)\; c_8) + 2\; c_9}{p^4}+...
\end{equation}

For $P=Q=R=(p,0)$;
\begin{equation}
[|G^{(4)}( P, Q, R)|]=-\frac {4/3}{p^6}-\frac {91/72}{p^4}+...,
\end{equation}
\begin{equation}
{\bf G}^{(4)}( P,Q,R)=-\frac {4/3}{p^6}+\frac {-17/9 + (40/3)\; c_7 + (40/3)\; c_8 - 8\; c_9}{p^4}+...
\end{equation}

Matching conditions therefore give three linear algebraic equations on the coefficients $c_7,\; c_8\; c_9$ and once again lead to the solution:
\begin{equation}
c_7 =c_8 = 0, \, c_9 = -5/64.
\end{equation}

\section{Conclusion}

To summarize we found a systematic way to perform RG of the decimation type in $d>1$ perturbatively. We have seen during our discussion that perturbative decimation RG has rather complicated structure including considerable number of extra contributions compared to more customary approaches (as momentum space RG) as well as some cumbersome numerical calculations. Here we would like to some of its uses. 

First of all, the formalism we propose here, being based on the decimation RG transformations, pssesses all the advantages of this type of RG. As we have seen, it operates with original fields only and does not require any (linear or nonlinear) transformations of variables. Besides, it preserves all the local relations including constraints. This in turn means that effective (coarse grained) theory will obey exactly the same local constraints as did original, and that no non-covariant terms will appear in the effective action. Among other applications, this opens a possibility to employ this formalism to study the critical phenomena, when strict control over symmetry properties of model becomes particularly important.

Another, compared to others \cite{Migdal-Kadanoff,Patkos,Swendsen}, useful feature of proposed formalism is its perturbative character. This can provide us with systematic method of calculations in asymptotically free models and, what is even more essential, with a way to do {\it controllable} approximations. Hopefully, this side of proposed formalism will make it applicable in situations when such a control is essential, as in above mentioned critical phenomena or in the recently proposed double strong-weak expansion approach \cite{Rosenstein}. It turns out that in asymptotically free theories there exist region in parameter space where both strong and weak coupling expansions are valid at the same time. Namely both the practical weak coupling $\alpha(g)=const\; g$ and strong coupling $\beta(g)=const/g$ expansion parameters are reasonably small. The "loop
factors" $1/(4\pi)^2$  in the practical weak coupling expansion 
parameter $\alpha(g)$ are partly responsible for this. In this scheme high frequency modes are integrated out perturbatively and the resulting effective action treated using strong coupling expansion.The symmetry preserving and controllable perturbative decimation technique is the most suitable tool for the first part of such calculations. Of course, to apply the method described here to the one dimensional models one should take into account correctly the nonperturbative instanton-like configurations, because in $d=1$ the perturbation theory can be ill defined (one such example we considered in Section 5). In higher dimensions, however, the relative contribution of the nonperturbative configurations becomes less significant.

We would like to stress also that the method described here unlike most of the other decimation (and exact RG in general) techniques, enables us to perform decimations not restricted to the simplest case $\eta=2$ only. The $\eta >2$ calculation just takes a bit more computer time.

\section{Appendix A. $d=1$ free fermion decimated action}

In this Appendix free Wilson fermions will be considered and decimated fermionic action in one dimension with $\eta =2$ will be derived via matching.  

The lattice action of the $d=1$ Wilson fermions has the form:
\begin{equation}
 S= \sum_{xy} \bar \psi_x \Delta_{xy} \psi_y,
 \end{equation}
where
$$
 \Delta_{xy} = \frac {1}{2}[\delta (x-y+1)-\delta (x-y-1)] + a m \delta (x-y)
$$
\begin{equation}
 - r [\delta (x-y+1) + \delta (x-y+1) -2 \delta (x-y)].
 \end{equation}
The Fourier transformed kernel is:
\begin{equation}
 \Delta (k) = i {\rm sin} (k a) + a m + 2 r {\rm sin}^2 \frac {a k }{2}.
 \end{equation}

Decimated propagator then reads:
\begin{equation}
 {\bf \Delta}^{-1} (K) = \sum_{n=1}^2  \frac{1}{i {\rm sin} \frac {K +2 \pi n}{2} + \frac{m}{2} + 2 r {\rm sin}^2 \frac {K+2 \pi n}{2}}.
\end{equation}

First of all, notice that at $r=0$ this expression does not have the correct form of the fermion propagator, and is instead 
\begin{equation}
 {\bf \Delta}^{-1} (K) =  \frac{m/2}{m^2/4 + {\rm sin}^2 \frac {K}{2}}.
\end{equation}
In particular, the corresponding effective action has no massless limit.
The reason of such a strange behaviour is that actually in massless fermion theory without the Wilson term there is no way to build an exponential generating functional for decimated theory and therefore there is no way to define an effective action. Moreover, one can check that in such a theory correlators between the even (or odd) cites vanish, so that $\eta=2$ decimation in this case leads to the complete loss of information.

In general case, the propagator takes the form:
\begin{equation}
 {\bf \Delta}^{-1} (K) =  \frac{m+2 r}{m^2/4+r m + 
(r^2+1) {\rm sin}^2 \frac {K}{2} +i r {\rm sin} K},
\end{equation}
that is the kernel of an effective action is
\begin{equation}
 {\bf \Delta} (K) =  \frac {r}{m+2 r} (i r {\rm sin} \;K + \frac {m(m+4 r)}{4 r} + 
(r+ \frac {1}{r}) {\rm sin}^2 \frac {K}{2}).
\end{equation}

Notice that for massless theory effective action after field rescaling has the same form with the new Wilson parameter
$$
r'=\frac {1}{2}(r+\frac{1}{r}).
$$
This means that the massless Wilson action in $d=1$ is a perfect action and has a fixed point $r=1_+$.

\section{Appendix B. Diagrammatic derivation of $d=1$ free boson decimated action}

Here we would like to reproduce the effective action in $d=1$ free boson model by means of the perturbation theory. Diagrammatics in this case consists of the internal propagator $G(k)$, external leg $\dbar (k)$ and the inverse decimated propagator ${\bf \Delta}(K)$:
$$
G(k)=\frac {a^2}{4 {\rm sin}^2 \frac {a k}{2}},\quad \dbar(k)=\frac {2}{a} {\rm cos}(a k),\quad {\bf \Delta}(K)=4 {\rm sin}^2 \frac {K}{2}.
$$

Diagrams contributing to the effective action are shown on Fig. 17 or 

\begin{picture}(5000,15000)(-10000,0)
\put(-4000,12700){$-\frac {2}{a}$}
\put(500,13000){\circle{500}}
\put(-500,13000){\circle{500}}
\put(3000,12700){$+$}
\put(5000,12700){$\frac {4}{a^2}$}
\drawline\photon[\E\REG](7000,13000)[3]
\put(\pfrontx,\pfronty){\circle{500}}
\drawline\fermion[\E\REG](\photonbackx,13000)[5000]
\drawline\photon[\E\REG](\fermionbackx,13000)[3]
\put(\photonbackx,\photonbacky){\circle{500}}
\put(-2000,7700){$-$}
\put(0,7700){$\frac {1}{a^2}$}
\drawline\photon[\E\REG](2500,8000)[3]
\put(\pfrontx,\pfronty){\circle{500}}
\drawline\fermion[\E\REG](\photonbackx,8000)[4000]
\THICKLINES
\drawline\fermion[\E\REG](\pbackx,8000)[3000]
\THINLINES
\put(\fermionfrontx,\fermionfronty){\circle{500}}
\drawline\fermion[\E\REG](\pbackx,8000)[4000]
\put(\fermionfrontx,\fermionfronty){\circle{500}}
\drawline\photon[\E\REG](\fermionbackx,8000)[3]
\put(\photonbackx,\photonbacky){\circle{500}}
\put(-3000,3000){Fig. 17. Diagrammatic representation of $d=1$ decimated action.}
\end{picture}
$$
S^{eff} = -\frac{1}{2} \int \phi_K \phi_{ -K} {\bf \Delta}(K)
$$
with 
\begin{equation}
{\bf \Delta}(K) =-\frac {2}{a} +\frac {4}{a^2}[| \frac {a^2 {\rm cos}^2 a K}{4 {\rm sin}^2 \frac {a K}{2}}|] - \frac {4}{a^2}[| \frac {a^2 {\rm cos} a K}{4 {\rm sin}^2 \frac {a K}{2}}|]^2 4 {\rm sin}^2 \frac {K}{2},
\end{equation}
where $[|f(k)|]$ denotes decimated expression: \
$$
[|f(K)|]=\frac {a}{2 \pi} \sum_X \int_{-\pi /a}^{\pi /a} e^{-i (K - p)X} f(p) d p =  \sum_{n=1}^{\frac {1}{a}} f(K+2 \pi n).
$$
The decimated blocs in an effective action are: 

\begin{picture}(10000,7000)(0,0)
\drawline\photon[\E\REG](2000,3000)[3]
\put(\photonfrontx,\photonfronty){\circle{500}}
\drawline\fermion[\E\REG](\photonbackx,3000)[5000]
\put(\fermionbackx,3000){\circle{500}}
\global\advance \pbackx by 2000
\put(\pbackx,\pbacky){$= [\vert \frac {a^2 {\rm cos}(a K)}{4 {\rm sin}^2 \frac {a K}{2}}\vert ]=\frac{1}{4 {\rm sin}^2 \frac {K}{2}} -\frac {a}{2}$,}
\end{picture}

\begin{picture}(10000,5000)(0,0)
\drawline\photon[\E\REG](2000,3000)[3]
\put(\photonfrontx,\photonfronty){\circle{500}}
\drawline\fermion[\E\REG](\photonbackx,3000)[5000]
\drawline\photon[\E\REG](\fermionbackx,3000)[3]
\put(\photonbackx,3000){\circle{500}}
\global\advance \pbackx by 2000
\put(\pbackx,\pbacky){$= [\vert \frac {a^2 {\rm cos}^2(a K)}{4 {\rm sin}^2 \frac {a K}{2}}\vert ]=\frac{1}{4 {\rm sin}^2 \frac {K}{2}} -\frac {a}{2}$.}
\end{picture}

Thus, for the kernel of an effective action we obtain an expression
\begin{equation}
{\bf \Delta}(K) =-\frac {2}{a} +\frac {4}{a^2}\left(\frac{1}{4 {\rm sin}^2 \frac {K}{2}} -\frac {a}{2}\right) - \frac {4}{a^2}\left(\frac{1}{4 {\rm sin}^2 \frac {K}{2}} -\frac {a}{2}\right)^2 4 {\rm sin}^2 \frac {K}{2} = 4 {\rm sin}^2 \frac {K}{2},
 \end{equation}
which clearly coincides with that obtained by the matching method.

\section{Appendix C. Quartic part of the $d=2$ $\sigma$-model effective action}

In this Appendix an expression for the quartic part of an effective action of $d=2$ $\sigma$-model will be given. This expression is necessary when the matching approach is employed.

Decimation technique is applied to the unconstrained variables, "pions" $\pi^i$ and therefore mainly gives us non-covariant quantities. On the other hand, an effective action is expressed in terms of constrained, covariant variables $S^a$. To reconstruct it from the non-covariant RG results, we need to re-express this effective action in terms of the fields $\pi^i$ and then its coefficients can be identified by simple comparison. 

The most general covariant effective action with up to four derivatives is given by an expression \cite{Simanzik}
$$
A^{eff}=g^{-1} \sum_X \left[ \frac {1}{2} S^a\Box S^a+c_5(\Box S^a)^2+(-\frac {1}{24} +c_6) \sum_\mu(\partial_\mu \partial_\mu^+ S^a)^2\right.
$$
\begin{equation}
\left.+c_7(S^a\Box S^a)^2+c_8\sum_\mu (S^a \partial_\mu \partial_\mu^+ S^a)^2+c_9\sum_{\mu \nu} \left (\frac {\partial_\mu + \partial_\mu^+}{2} S^a . \frac {\partial_\mu + \partial_\mu^+}{2} S^a\right)^2 \right ].
\end{equation}
Here $S^a$ ($a=1,2,..,N$) are $O(N)$ vectors on lattice normalized to unity: $S^2=1$, and we follow notations of \cite{Simanzik} with lattice spacing $A=1$.

To reconstruct coefficients $c_5,..,c_9$, one can solve the constraint:
$$
S^a=(\sqrt{g} \pi^i,\sqrt {1-g (\pi^i)^2}), i=1,..,N-1,
$$
 and expand this action in terms of "pions". To the fourth order in $\pi^i$ and up to fourth derivatives, actions is:
$$
A^{eff}=A^{(2)}+A^{(4)}
$$
with quadratic part 
\begin{equation}
A^{(2)}= \sum_X \left[ \frac {1}{2} \pi^i\Box \pi^i
+c_5(\Box \pi^i)^2+(-\frac {1}{24} +c_6) \sum_\mu(\partial_\mu \partial_\mu^+ \pi^i)^2\right]
\end{equation}
and quartic part
$$
A^{(4)}=g \sum_X \left[ \frac {1}{8} \pi^2\Box \pi^2
+ \frac {1}{4}c_5(\Box \pi^2)^2+\frac {1}{4}(-\frac {1}{24} +c_6) \sum_\mu(\partial_\mu \partial_\mu^+ \pi^2)^2 \right.
$$
\begin{equation}
\left. + c_7(\pi^i\Box \pi^i)^2+ c_8\sum_\mu (\pi^i \partial_\mu \partial_\mu^+ \pi^i)^2+ c_9\sum_{\mu \nu} \left (\frac {\partial_\mu + \partial_\mu^+}{2} \pi^i . \frac {\partial_\nu + \partial_\nu^+}{2} \pi^i \right )^2 \right].
\end{equation}

\section{Appendix D. One loop contributions to the four derivatives terms of the effective action}

Here the leading in $1/N$ expansion one loop diagrams contributing to the four derivatives terms of the effective action are shown. All the diagrams can be divided into four groups: diagrams with all the external points coinciding (Fig. 18(a), 18(b)); with two pairs of the coinciding points(Fig. 18(c)); with two coinciding and two different points (Fig. 19) and with all different external points (Fig. 20). Diagrams    

\begin{picture}(45000,17000)(-3000,-7500)

\put(8500,500){\circle{500}}
\put(8500,-500){\circle{500}}
\put(7500,-500){\circle{500}}
\put(7500,500){\circle{500}}
\drawline\photon[\N\REG](8000,1000)[2]
\put(8000,5000){\circle{5000}}
\put(8000,-2500){($a$)}

\put(15500,2500){\circle{500}}
\put(15500,1500){\circle{500}}
\drawline\photon[\E\REG](16000,2000)[2]
\put(20300,2000){\circle{5000}}
\drawline\photon[\E\REG](22500,2000)[2]
\put(25000,2500){\circle{500}}
\put(25000,1500){\circle{500}}
\put(19500,-2000){($b$)}

\put(-2000,-5000){Fig. 18. Contributions with 4 [figure (a)] and 2+2 (b) coinciding points.}
\end{picture}
from different groups give different analytical expressions even for $d=1$,
 where diagrams inside each group are proportional to each other. Thus, these groups can be calculated independently. This can provide us with

\begin{picture}(45000,22000)(0,-17000)
\put(-500,-3500){\circle{500}}
\put(-500,-4500){\circle{500}}
\drawline\photon[\E\REG](0,-4000)[3]
\drawline\fermion[\NE\REG](\pbackx,\pbacky)[5000]
\drawline\gluon[\N\FLIPPED](\pmidx,\pmidy)[2]
\global\Yone \pbacky
\global\advance \Yone by 1000
\put(\pbackx,\Yone){\circle{2000}}
\drawline\photon[\NE\REG](\fermionbackx,\fermionbacky)[2]
\put(\pbackx,\pbacky){\circle{500}}
\drawline\fermion[\SE\REG](\fermionfrontx,\fermionfronty)[5000]
\drawline\photon[\SE\REG](\pbackx,\pbacky)[2]
\put(\pbackx,\pbacky){\circle{500}}
\put(2500,-13000){($a$)}

\put(12500,-3500){\circle{500}}
\put(12500,-4500){\circle{500}}
\drawline\scalar[\S\REG](16500,-2000)[2]
\global \Yfour \pfronty
\global\advance \Yfour by 1000
\global\advance \scalarbackx by 200
\put(\pbackx,\Yfour){\circle{2000}}
\drawline\photon[\E\REG](13000,\pmidy)[3]
\drawline\fermion[\NE\REG](\scalarbackx,\scalarbacky)[4000]
\drawline\photon[\NE\REG](\pbackx,\pbacky)[2]
\put(\pbackx,\pbacky){\circle{500}}
\drawline\fermion[\SE\REG](\fermionfrontx,\fermionfronty)[4000]
\drawline\photon[\SE\REG](\pbackx,\pbacky)[2]
\put(\pbackx,\pbacky){\circle{500}}
\put(14500,-13000){($b$)}

\put(25500,-3500){\circle{500}}
\put(25500,-4500){\circle{500}}
\drawline\photon[\E\REG](26000,-4000)[2]
\put(30300,-4000){\circle{5000}}
\drawline\gluon[\E\REG](32500,-4000)[2]
\drawline\fermion[\SE\REG](\pbackx,\pbacky)[4000]
\drawline\photon[\SE\REG](\pbackx,\pbacky)[2]
\put(\pbackx,\pbacky){\circle{500}}
\drawline\fermion[\NE\REG](\gluonbackx,\gluonbacky)[4000]
\drawline\photon[\NE\REG](\pbackx,\pbacky)[2]
\put(\pbackx,\pbacky){\circle{500}}
\put(29500,-13000){($c$)}

\put(3000,-15000){Fig. 19. Contributions of the 2+1+1 type.}

\end{picture}
an additional consistency checks (cancellations of the negative powers of the external momenta inside the groups).

\begin{picture}(45000,33000)(-3000,-30000)

\drawline\gluon[\E\REG](3000,-3000)[3]
\global\advance \pmidy by -8000
\global\advance \pmidx by -500
\put(\pmidx,\pmidy){($a$)}
\drawline\fermion[\NW\REG](\pfrontx,\pfronty)[3000]
\drawline\photon[\NW\REG](\pbackx,\pbacky)[2]
\put(\pbackx,\pbacky){\circle{500}}
\drawline\fermion[\SW\REG](\fermionfrontx,\fermionfronty)[3000]
\drawline\photon[\SW\REG](\pbackx,\pbacky)[2]
\put(\pbackx,\pbacky){\circle{500}} 
\drawline\fermion[\NE\REG](\gluonbackx,\gluonbacky)[3000]
\drawline\gluon[\N\FLIPPED](\pmidx,\pmidy)[2]
\global\advance \pbacky by 1000
\put(\pbackx,\pbacky){\circle{2000}}
\drawline\photon[\NE\REG](\fermionbackx,\fermionbacky)[2]
\put(\pbackx,\pbacky){\circle{500}} 
\drawline\fermion[\SE\REG](\fermionfrontx,\fermionfronty)[3000]
\drawline\photon[\SE\REG](\pbackx,\pbacky)[2]
\put(\pbackx,\pbacky){\circle{500}}

\drawline\gluon[\E\REG](21000,-3000)[3]
\global\Xthree \pbackx
\global\advance \Xthree by 500
\global\advance \pmidy by -8000
\global\advance \pmidx by -500
\put(\pmidx,\pmidy){($b$)}
\drawline\fermion[\NW\REG](\pfrontx,\pfronty)[3000]
\drawline\photon[\NW\REG](\pbackx,\pbacky)[2]
\put(\pbackx,\pbacky){\circle{500}}
\drawline\fermion[\SW\REG](\fermionfrontx,\fermionfronty)[3000]
\drawline\photon[\SW\REG](\pbackx,\pbacky)[2]
\put(\pbackx,\pbacky){\circle{500}}
\drawline\scalar[\S\REG](\Xthree,-1000)[2]
\global\advance\scalarbackx by 200
\global\advance\scalarfronty by 1000
\put(\scalarfrontx,\scalarfronty){\circle{2000}}
\drawline\fermion[\NE\REG](\scalarbackx,\scalarbacky)[3000]
\drawline\photon[\NE\REG](\fermionbackx,\fermionbacky)[2]
\put(\pbackx,\pbacky){\circle{500}} 
\drawline\fermion[\SE\REG](\fermionfrontx,\fermionfronty)[3000]
\drawline\photon[\SE\REG](\pbackx,\pbacky)[2]
\put(\pbackx,\pbacky){\circle{500}}

\drawline\scalar[\E\REG](3000,-18000)[3]
\global\Xfive \pmidx
\global\Yfive \pmidy
\global\advance \Yfive by -6000
\global\advance \Xfive by -500
\put(\Xfive,\Yfive){($c$)}
\drawline\gluon[\N\REG](\pmidx,\pmidy)[2]
\global\advance \pbacky by 1000
\put(\pbackx,\pbacky){\circle{2000}}
\drawline\fermion[\NW\REG](\scalarfrontx,\scalarfronty)[3000]
\drawline\photon[\NW\REG](\pbackx,\pbacky)[2]
\put(\pbackx,\pbacky){\circle{500}}
\drawline\fermion[\SW\REG](\fermionfrontx,\fermionfronty)[3000]
\drawline\photon[\SW\REG](\pbackx,\pbacky)[2]
\put(\pbackx,\pbacky){\circle{500}} 
\drawline\fermion[\NE\REG](\scalarbackx,\scalarbacky)[3000]
\drawline\photon[\NE\REG](\pbackx,\pbacky)[2]
\put(\pbackx,\pbacky){\circle{500}} 
\drawline\fermion[\SE\REG](\fermionfrontx,\fermionfronty)[3000]
\drawline\photon[\SE\REG](\pbackx,\pbacky)[2]
\put(\pbackx,\pbacky){\circle{500}}

\drawline\gluon[\E\REG](20000,-18000)[2]
\drawline\fermion[\SW\REG](\pfrontx,\pfronty)[3000]
\drawline\photon[\SW\REG](\pbackx,\pbacky)[2]
\put(\pbackx,\pbacky){\circle{500}}
\drawline\fermion[\NW\REG](\gluonfrontx,\gluonfronty)[3000]
\drawline\photon[\NW\REG](\pbackx,\pbacky)[2]
\put(\pbackx,\pbacky){\circle{500}}
\put(24300,-18000){\circle{5000}}
\drawline\gluon[\E\REG](26500,-18000)[2]
\drawline\fermion[\SE\REG](\pbackx,\pbacky)[3000]
\drawline\photon[\SE\REG](\pbackx,\pbacky)[2]
\put(\pbackx,\pbacky){\circle{500}}
\drawline\fermion[\NE\REG](\gluonbackx,\gluonbacky)[3000]
\drawline\photon[\NE\REG](\pbackx,\pbacky)[2]
\put(\pbackx,\pbacky){\circle{500}}
\put(23500,-24000){($d$)}

\put(0,-27000){Fig. 20. Contributions of $1+1+1+1$ type.}

\end{picture}

\end{document}